\documentclass[journal]{IEEEtran}

\usepackage{cite}
\usepackage{colortbl}  % for \rowcolor
\usepackage{amsmath,amssymb}
\usepackage{graphicx}
\usepackage{multirow}
\usepackage{url}
\usepackage[caption=false,font=footnotesize]{subfig}
\usepackage{booktabs}
\usepackage{array}
\newcolumntype{C}[1]{>{\centering\arraybackslash}p{#1}} % centered p-column
\usepackage{threeparttable}
\usepackage{balance}

\begin{document}
\title{Path-Specific Geometric--Stochastic Channel Model for Scattering from ISAC Targets:\\
A Unified Measurement-Based Characterization of Robots, Humans, and
Automobiles at 28~GHz}

\author{Neeraj~Varshney,~\IEEEmembership{Senior Member,~IEEE},
Jelena~Senic,
Jack~Chuang,
Samuel~Berweger,~\IEEEmembership{Member,~IEEE},
Steve~Blandino,
Jian~Wang,
Camillo~Gentile,~\IEEEmembership{Member,~IEEE},
and~Nada~Golmie,~\IEEEmembership{Fellow,~IEEE}%
\thanks{All authors are with the Communications Technology Laboratory, National Institute of Standards and Technology (NIST). N. Varshney, J. Chuang, S. Blandino, J. Wang, C. Gentile, and N. Golmie are located in Gaithersburg, MD 20899, USA. J. Senic and S. Berweger are located in Boulder, CO 80305, USA. N. Varshney and S. Blandino are also with Prometheus Computing LLC, Bethesda, MD 20814, USA. C.~Gentile is the corresponding author (e-mail: camillo.gentile@nist.gov).}
}

% Running head
\markboth{Submitted to IEEE Transactions on Antennas and Propagation}%
{Varshney \MakeLowercase{\textit{et al.}}: Multi-Scatterer Target Models for ISAC}

\maketitle

\begin{abstract}
This paper develops a channel model for scattering from a broad
class of integrated sensing and communication (ISAC) targets—robots, humans, and automobiles—based
on a unified measurement campaign at 28~GHz. Nine targets are characterized from numerous transmitter--receiver
configurations per target, spanning full bistatic and aspect angles and thus
covering backward to forward scattering. Repeated
acquisitions are collected for each configuration to capture target motion
and dynamics, yielding nearly 117\,000 acquisitions in total. Measurements
are collected with a 2~GHz-wide channel sounder equipped with a
256-element phased-array antenna, followed by digital beamforming
and super-resolution processing for reliable extraction
of multiple paths in delay and 2D angle, with average estimation errors of 0.1~ns and 0.2$^\circ$  respectively. 
From these measurements, we reduce a \emph{path-specific} channel representation in which the scattering is modeled as a cluster of paths whose center is defined by the bistatic specular \emph{geometry}; the cluster shape is described \emph{stochastically}. To ensure physically grounded spatio--temporal consistency,
extracted paths are mapped to incident points on the target, each with an associated radar cross section (RCS) that is compatible with
the 3GPP TR~38.901 ISAC channel model, and are subsequently mapped back to path properties for channel synthesis. The
resulting statistical distributions are well behaved, physically
interpretable, and consistent across targets, geometries, and motion.
Comparison between measured and synthesized channel realizations
demonstrates close agreement in pathloss trends and fluctuation
behavior. Together, these results yield a high-fidelity,
measurement-derived channel representation suitable for ISAC
simulation, target modeling, and system design.
\end{abstract}

\section{Introduction}
\IEEEPARstart{I}{ntegrated} sensing and communication (ISAC) systems envisioned
for future 6G networks require accurate characterization of how dynamic
targets—such as unmanned aerial vehicles (UAVs), robots, humans, and automobiles—influence the radio-frequency (RF)
propagation channel. Beyond link-level performance, ISAC applications rely
on target-dependent scattering behavior to enable sensing tasks including
detection, identification, classification, and tracking~\cite{wang2022integrated}.
Recent multimodal ISAC frameworks further combine RF sensing with
modalities such as cameras to improve beam alignment, tracking, and
resource allocation in highly mobile networks~\cite{fan2026heterogeneous,fan2026optimizing}. Achieving these
capabilities requires not only knowledge of aggregate scattering strength,
but also an understanding of how scattering geometry evolves with target
motion, orientation, and structure. In particular, sensing performance
depends on how target-induced multipath is organized spatially and how
this structure evolves over time. This need has motivated extensive
measurement-based investigations of target radar cross section (RCS) across
a wide range of frequencies, target classes, and experimental setups.

A substantial body of recent work reports measured RCS characteristics of
individual targets, including UAVs at millimeter-wave (mmWave) bands
\cite{semkin2020analyzing,yuan2024experimental}, at microwave bands
\cite{yuan2024experimental,semkin2021drone}, and for detection
performance \cite{rosamilia2022radar}; road vehicles and
infrastructure from 23~GHz to 27~GHz \cite{schipper2011rcs}; automotive
targets at 94~GHz \cite{kamel2017rcs}; and vehicular ISAC scenarios in
which cars and trucks act as both reflectors and blockers
\cite{nataraja2025cars}. Related investigations focus on human targets,
providing human signatures for ISAC radar and breath-activity monitoring
\cite{piuzzi2015complex}, wideband data for behavior classification
\cite{fortuny2013radar}, near-field pedestrian RCS simulation and
validation at 23--28~GHz \cite{manfredi2020efficient}, and
pedestrian-oriented ISAC channel measurements
\cite{ebrahimizadeh2024millimeter}. Additional studies characterize object
and vehicle scattering using frequency-modulated continuous-wave (FMCW)
radars \cite{lee2016radar} and examine RCS patterns for automotive radar
at 77~GHz~\cite{he2022channel}. While these efforts deliver valuable
empirical insight, they employ different hardware platforms, frequencies,
geometries, calibration procedures, and extraction methods, which makes it
difficult to unify results or derive a consistent target model applicable
across ISAC scenarios.

To generalize measured behavior, several works fit statistical distributions
such as Lognormal, Nakagami, or Rician models to RCS data in order to capture
inter-aspect variability and large-scale scattering effects. Examples
include multi-band UAV RCS statistics \cite{das2025monostatic}, statistical
characterization of UAV and robot signatures
\cite{waqar2025statistical}, 3GPP-compliant RCS characterization of
indoor-factory UAV, robotic-arm, and AGV targets \cite{azim20253gpp},
and analyses of angular and frequency dependence motivated by ISAC
applications \cite{fenollosa2024frequency}. These approaches improve
descriptive understanding of target scattering, but they remain
fundamentally scalar in nature, providing distributions of total RCS rather
than models tied to resolvable propagation paths. As a result, they do not
support channel realizations or sensing algorithms that depend on path-level
geometry or spatial consistency.

Other studies move toward more structured abstractions by combining
measurements with simplified RCS models or ray-based simulations. Several
works analyze RCS frequency dependence to inform ISAC system design
\cite{zhang2025unified}, propose unified target-level RCS factors for link
evaluation \cite{zhang2025research}, or investigate monostatic UAV RCS for
ISAC channel development \cite{yuan2024experimental}. Hybrid approaches
combine measurements with raytracing to tune material parameters or
validate aggregate scattering behavior
\cite{luo2024channel,zhang2023channel}. While these models introduce useful
structure, validation is typically performed on the total received power
summed over all paths, rather than on individual scattering contributions.
Because most measurement systems lack sufficient delay and angular
resolution, the relationship between target geometry, bistatic
configuration, and path-specific scattering behavior remains largely
unmodeled.

In parallel, ISAC channel modeling has advanced rapidly, with several
frameworks extending traditional communication models to incorporate
bistatic geometries and target-induced scattering
\cite{ghosh2025unified,wei2024integrated}. Recent models introduce segmented
or concatenated ISAC channels that decouple background and target components
\cite{jiang2025segmented}. This separation is also consistent with the
3GPP TR~38.901 ISAC channel model~\cite{3gpp38901}, in which target
channel(s) and the background channel are generated separately and then
combined. Other models propose multi-scattering-point abstractions
\cite{yang2025isac,ye2024general}, formulate general ISAC channel structures
combining forward and backward scattering \cite{yang2023novel}, and embed
target scattering into hybrid geometric--stochastic models
\cite{zhang2022general}. Despite this progress, most existing models rely on
assumed target representations rather than measurement-resolved path data.
Moreover, multipoint abstractions often collapse to a single equivalent
scatterer after antenna integration, limiting their ability to preserve the
geometric detail required for target identification and classification.
Consequently, validation is typically restricted to aggregate channel
metrics, leaving path-specific scattering structure unverified.

This paper addresses these limitations by presenting a unified measurement
campaign coupled with a path-specific scattering model for representative ISAC
targets, including robots, humans, and automobiles. All targets are measured using the same 28~GHz channel sounder under consistent geometry, calibration, synchronization, and
processing, with multiple transmitter (TX)--receiver (RX) configurations
per target selected to systematically span full bistatic and aspect
angles, thereby covering backward to forward scattering. A 256-element phased-array receiver enables
high-resolution digital beamforming and super-resolution path extraction in
post-processing, allowing reliable recovery of multiple paths
even for compact targets such as UAVs and robotic arms. Measurement data presented in this paper is available at~\cite{nist_nextg_repository}. The measurements
reveal that target scattering is naturally organized as clusters of
resolvable paths centered around the specular reflection point.
From these observations, we derive a path-specific channel model that maps
extracted paths to incident points on the target, providing a common
geometric basis for spatial--temporal consistency and incident-point-specific
RCS characterization.
%By interpreting extracted paths through a geometric target
%representation, the proposed approach ensures spatial and temporal
%consistency across motion while preserving physically grounded
%scattering structure. The resulting model remains valid across diverse
%targets and configurations and naturally reduces to an effective
%single-scatterer abstraction when integrated with antenna patterns at long
%ranges.
The main contributions of this paper are as follows:
\begin{enumerate}
\item \textit{Unified multi-target measurement campaign:} A single 28~GHz
ISAC channel-sounding platform is used to measure a total of nine targets--robots, humans, and automobiles. Numerous TX--RX configurations
per target are selected to span full bistatic and aspect angles, and repeated acquisitions per configuration to capture target motion and dynamics, yielding nearly 117\,000 acquisitions in total.

\item \textit{Super-resolution path extraction:} A 256-element phased-array
RX with digital (switched) beamforming combined with a 2~GHz probing signal and
super-resolution processing enables reliable extraction of multiple paths even from compact targets, achieving average estimation errors of approximately $0.1$~ns in delay and $0.2^\circ$ in angle.

\item \textit{Path-specific geometric--stochastic channel model:}
A measurement-derived geometric--stochastic model is developed in which
target scattering is represented as clusters of paths anchored at the
specular reflection point. By mapping extracted paths to incident points on
the target, the model couples path geometry with incident-point-specific RCS
statistics while ensuring spatial--temporal consistency.
\end{enumerate}

The remainder of this paper is organized as follows.
Section~\ref{sec:Measurement Campaign} describes the measurement campaign.
Section~\ref{sec:sounder} presents the 28~GHz ISAC measurement system, including the path extraction and path-to-incident-point mapping pipeline.
Sections~\ref{sec:cm_geom} and~\ref{sec:cm_rcs} then develop the two coupled components of the proposed path-specific channel model.
Section~\ref{sec:cm_geom} defines the geometric component, in which target scattering is organized as a cluster of resolvable paths anchored at the specular reflection point, with cluster shape described stochastically.
Section~\ref{sec:cm_rcs} introduces the RCS component, which follows a 3GPP-compliant formulation and assigns incident-point-specific scattering strengths within the same incident-point representation.
Section~\ref{sec:channel_realization} combines these geometric and RCS components to generate example channel realizations and validate the model.
Section~\ref{sec:conclusion} concludes the paper.

% ============================================================
% TABLE + FIGURE TOGETHER (table on top, figure on bottom)
% ============================================================

\begin{table*}[!t]
\centering
\caption{Measurement Campaign Summary}
\label{tab:meascamp}
\renewcommand{\arraystretch}{1.15}
\setlength{\tabcolsep}{4.5pt}
% ---------------- TABLE ----------------
\begin{threeparttable}
\begin{tabular}{l c c c c c c c c}
\hline
Target &
\shortstack{TX--target\\standoff\\{[m]}} &
\shortstack{RX--target\\standoff\\{[m]}} &
\shortstack{TX/RX\\height\\{[m]}} &
\shortstack{No. of\\TX--RX\\cfg.} &
\shortstack{No. of\\acq. per\\TX--RX cfg.} &
\shortstack{Per-acq.\\displacement\\{[mm]}} &
\shortstack{Bistatic\\angle\\range [$^\circ$]} &
\shortstack{Aspect\\angle\\range [$^\circ$]} \\
\hline

UAV\textsuperscript{a} &
4.0 & 4.0 &
1.33 / 1.31 &
9 &
37 &
52 &
0--180 &
0--180 \\

Robotic Arm &
3.0 & 3.0 &
1.33 / 1.31 &
9 &
2000 &
0.52 &
0--180 &
35--140 \\

AMR  &
4.0 & 4.0 &
0.95 / 1.31 &
9 &
1860 &
2.15 &
0--180 &
0--180 \\

Humans (3) &
1.5 & 2.5 &
1.55 / 1.53 &
9 &
2000 &
1 &
0--180 &
5--180 \\

Automobiles (3) &
5.5--7.7 & 4.0--6.2 &
1.33 / 1.31 &
8 &
$\sim$1160 &
18.1 &
0--180 &
0--180 \\
\hline
\end{tabular}

\begin{tablenotes}[flushleft]
\footnotesize
\item[a] Since the UAV motion was purely rotational, the displacement corresponds to the arc length of a $5^\circ$ in-place rotation at an effective radius of 0.6 m ($52$ mm).
\end{tablenotes}
\end{threeparttable}

\vspace{0.6em}

% ---------------- FIGURE (INSIDE SAME TABLE*) ----------------
\begin{minipage}{\textwidth}
\centering
\renewcommand{\arraystretch}{0}

% --- size controls ---
\newcommand{\colWsmall}{0.28\textwidth}
\newcommand{\colWcar}{0.36\textwidth}
\newcommand{\rowH}{0.16\textheight}
\newcommand{\carW}{0.35\textwidth}

% --- spacing controls ---
\newcommand{\colgap}{3pt}
\newcommand{\rowgap}{1pt}

% --- build subfloats in COLUMN-MAJOR order so labels match caption ---
\newsavebox{\boxA}\newsavebox{\boxB}\newsavebox{\boxC} % robots: (a)--(c)
\newsavebox{\boxD}\newsavebox{\boxE}\newsavebox{\boxF} % humans: (d)--(f)
\newsavebox{\boxG}\newsavebox{\boxH}\newsavebox{\boxI} % autos: (g)--(i)

\sbox{\boxA}{\subfloat[]{\includegraphics[height=\rowH,keepaspectratio]{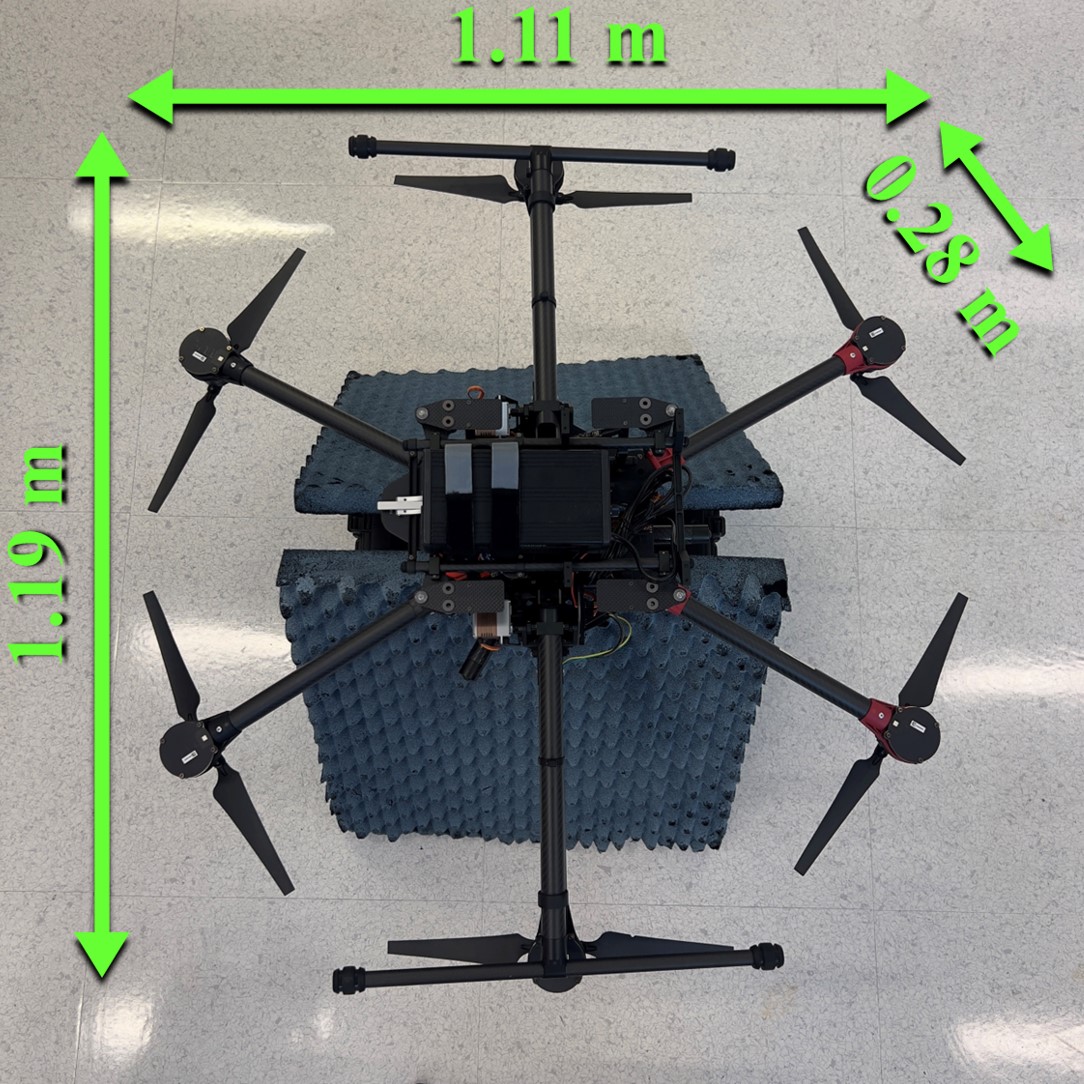}}}
\sbox{\boxB}{\subfloat[]{\includegraphics[height=\rowH,keepaspectratio]{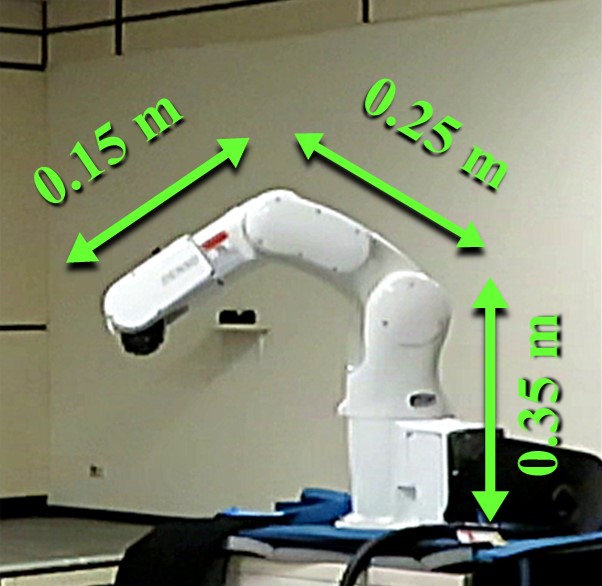}}}
\sbox{\boxC}{\subfloat[]{\includegraphics[height=\rowH,keepaspectratio,clip,trim=8mm 7mm 8mm 12mm]{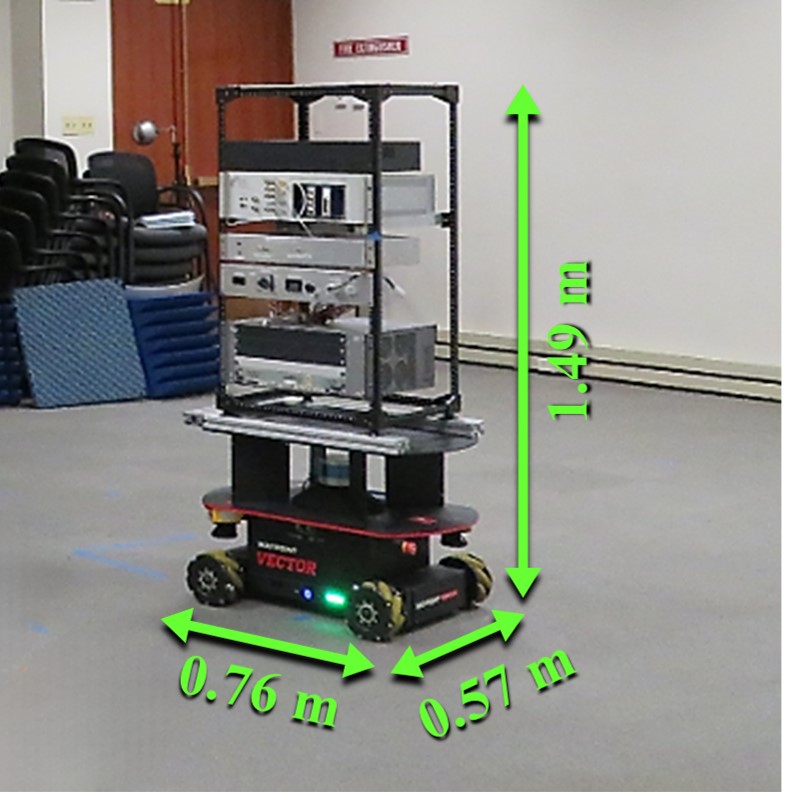}}}

\sbox{\boxD}{\subfloat[]{\includegraphics[height=\rowH,keepaspectratio]{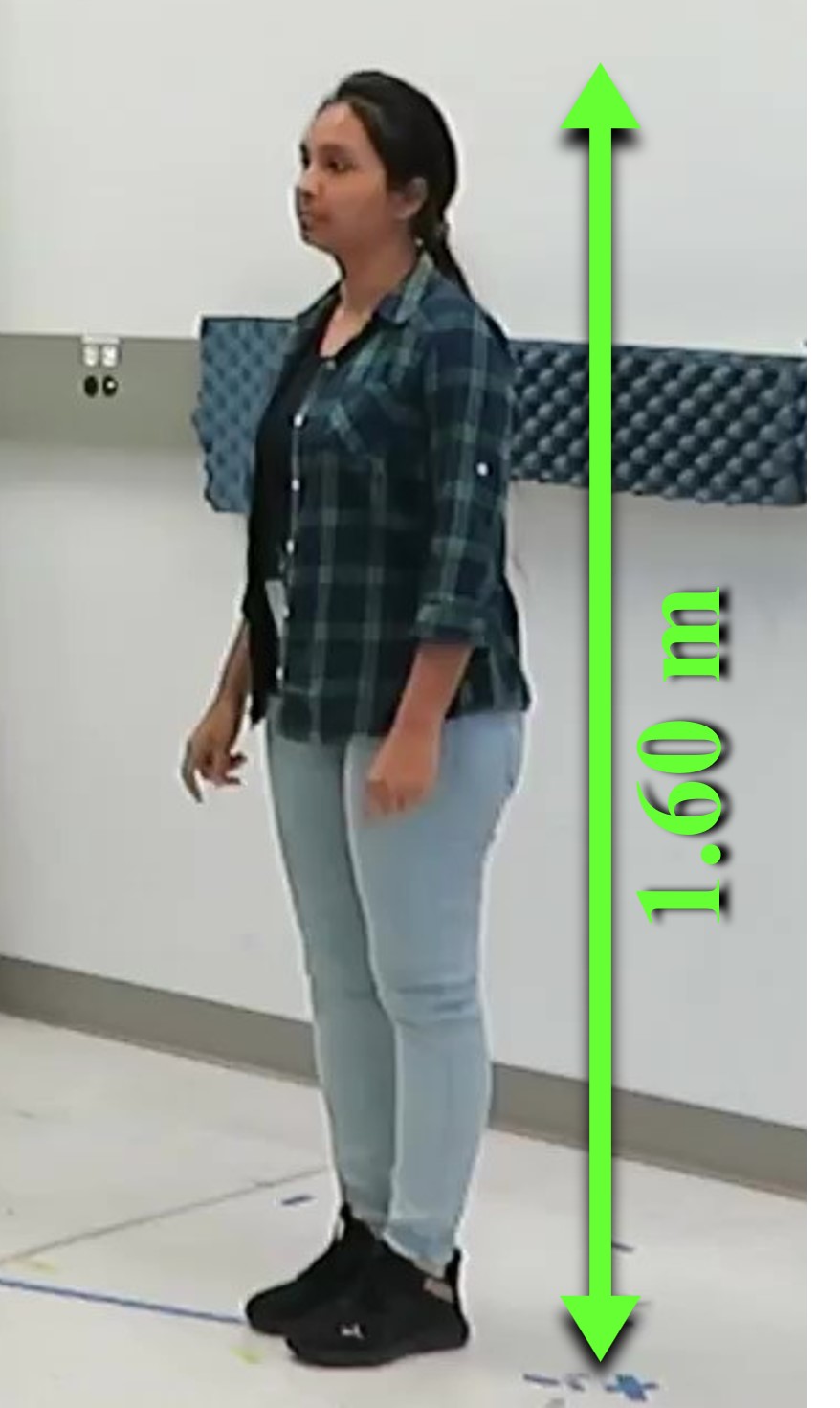}}}
\sbox{\boxE}{\subfloat[]{\includegraphics[height=\rowH,keepaspectratio]{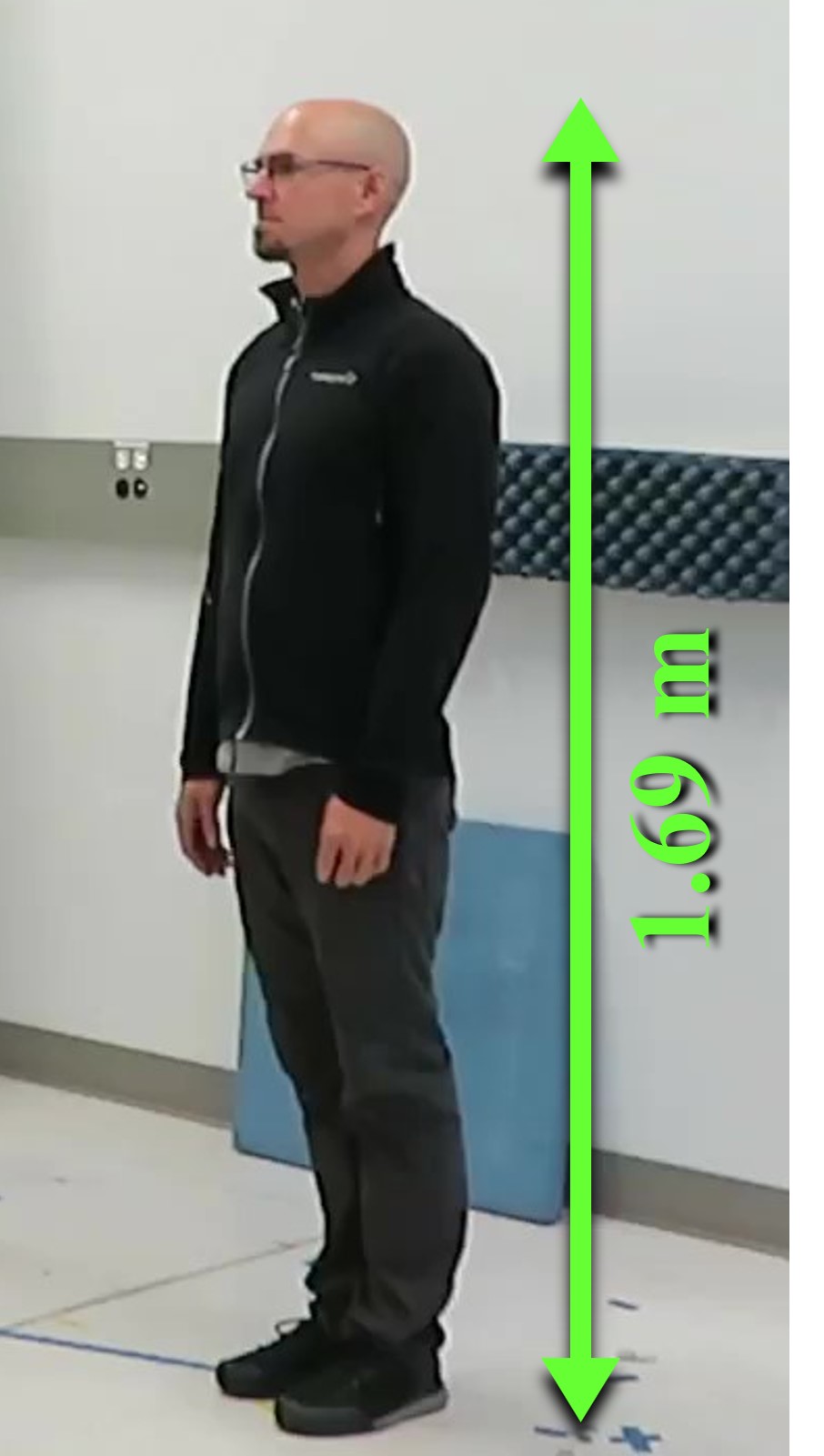}}}
\sbox{\boxF}{\subfloat[]{\includegraphics[height=\rowH,keepaspectratio]{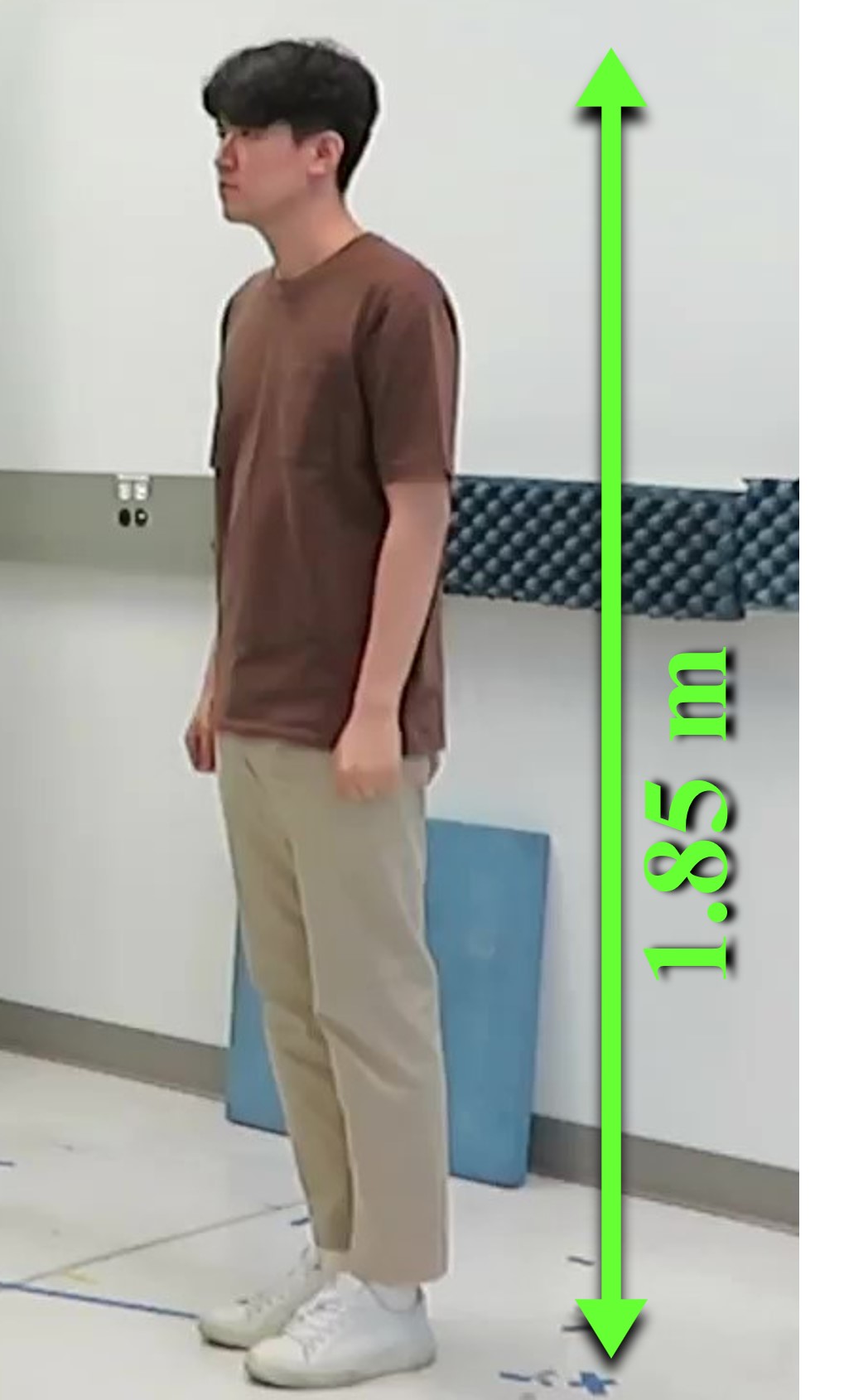}}}

\sbox{\boxG}{\subfloat[]{\includegraphics[width=\carW,keepaspectratio]{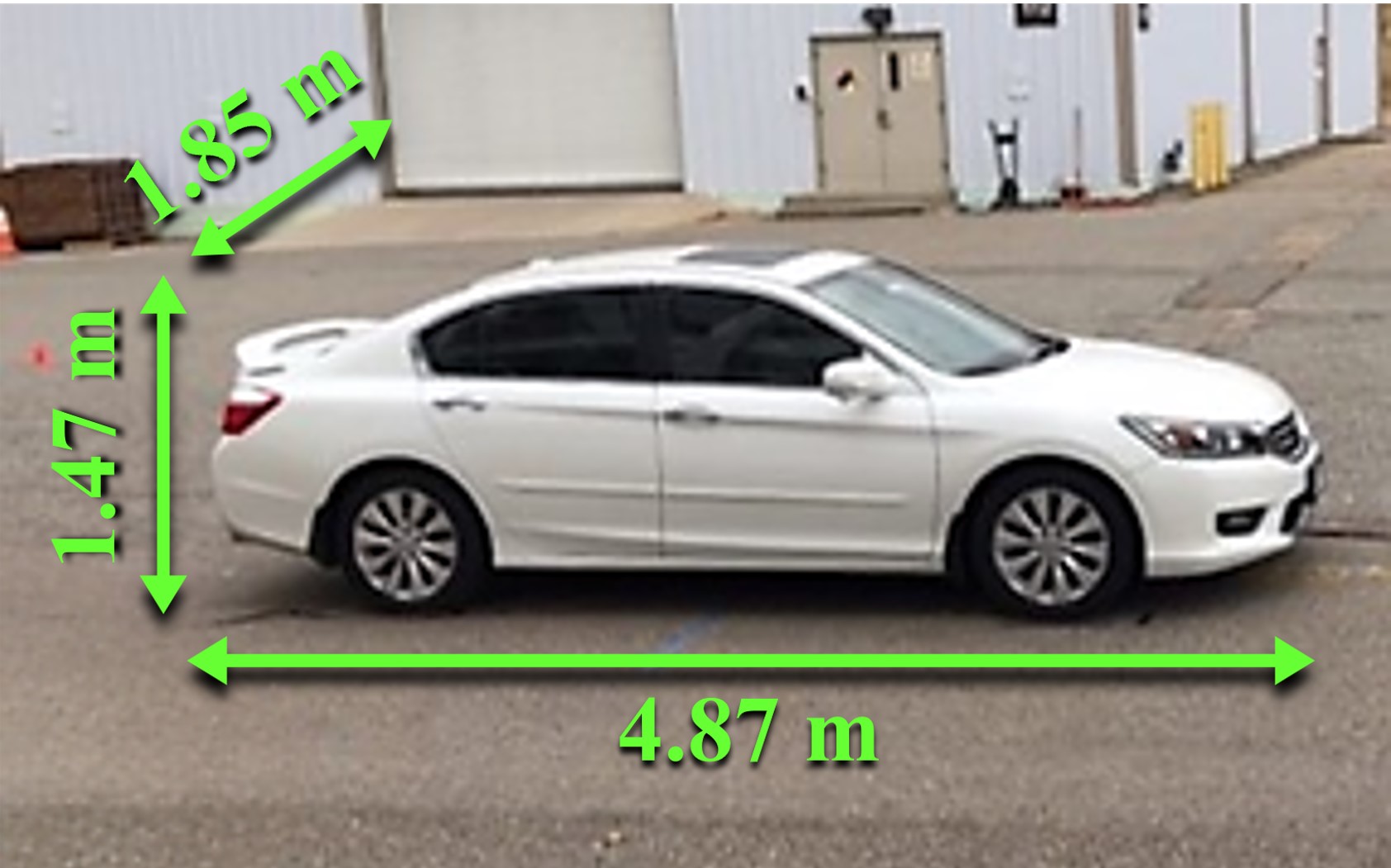}}}
\sbox{\boxH}{\subfloat[]{\includegraphics[height=\rowH,keepaspectratio]{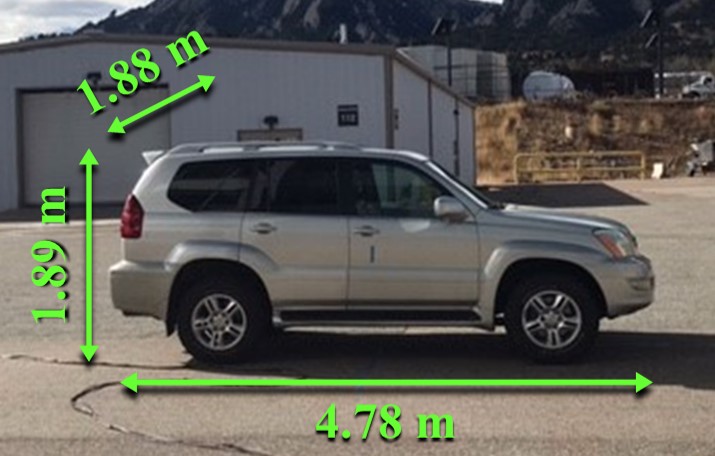}}}
\sbox{\boxI}{\subfloat[]{\includegraphics[height=\rowH,keepaspectratio]{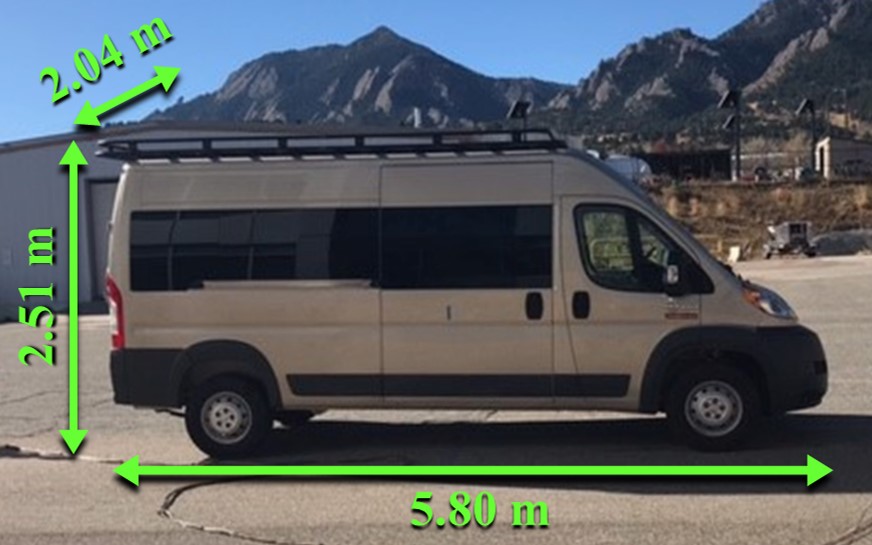}}}

% --- place the boxes in the SAME 3x3 visual layout as before ---
\begin{tabular}{@{}C{\colWsmall}@{\hspace{\colgap}}C{\colWsmall}@{\hspace{\colgap}}C{\colWcar}@{}}

% -------- Row 1 --------
\usebox{\boxA} & \usebox{\boxD} & \usebox{\boxG}
\\[\rowgap]

% -------- Row 2 --------
\usebox{\boxB} & \usebox{\boxE} & \usebox{\boxH}
\\[\rowgap]

% -------- Row 3 --------
\usebox{\boxC} & \usebox{\boxF} & \usebox{\boxI}

\end{tabular}

\refstepcounter{figure}
\vspace{0.35em}
\raggedright
{\scriptsize Fig.~\thefigure.
(a)--(c) Robots: UAV, robotic arm, and autonomous mobile robot (AMR).
(d)--(f) Human subjects.
(g)--(i) Automobiles: sedan, SUV, and van.}
\label{fig:targets9}

\end{minipage}
\end{table*}

\section{Measurement Campaign}
\label{sec:Measurement Campaign}
A unified measurement campaign was conducted on nine targets: a UAV, a robotic arm, an autonomous mobile robot (AMR),
three human subjects\footnote{All measurements involving human subjects were conducted in accordance with applicable National Institute of Standards and
Technology (NIST) guidelines. Informed consent was obtained from all participants prior to data collection.}, and three automobiles. Measurement data presented in this paper is available at~\cite{nist_nextg_repository}. Photographs\footnote{Certain commercial equipment, instruments, or materials
are illustrated in this paper in order to adequately specify the
experimental procedure. Such identification is not intended to imply
recommendation or endorsement by NIST, nor is it intended to imply that the materials or equipment
identified are necessarily the best available for the purpose.} of the targets
are shown in Fig.~\ref{fig:targets9}. The campaign comprised an ensemble of separate, target-specific variants that spanned different geometric sizes and material properties. In this way, target dimensions are represented
through the geometric model, while material-dependent scattering effects
are represented empirically through the fitted stochastic parameters
rather than through explicit material constants. This approach supports
a compact geometric--stochastic target-scatter model, rather than a
high-fidelity raytracing or electromagnetic-solver solution.

The objective was to characterize scattering from each target as a function of the aspect angle
$\phi$ and bistatic angle $\beta$, consistent with the 3GPP TR~38.901 channel model~\cite{3gpp38901}, in which target-specific scattering behavior is parameterized geometrically to support ISAC target-channel and RCS modeling.
To this end, two bistatic acquisition configurations were employed.
For the UAV, robotic arm, AMR, and humans, the RX was held stationary while the TX was repositioned
at nine azimuthal locations spaced $22.5^\circ$ apart on a semicircle
surrounding the target (the complementary
semicircle was not measured, exploiting the target bilateral symmetry).
For each TX position, the targets were set in motion
to capture dynamic scattering behavior as well as induce controlled variations in $\phi$ and $\beta$.
By contrast, the automobiles were kept stationary and the RX was repositioned at eight azimuthal locations
spaced $45^\circ$ apart on a circle surrounding the target. For each RX position,
the TX was mounted on a mobile platform and scanned only a semicircle around the target
(again exploiting its bilateral symmetry). This configuration
enabled precise and repeatable control of $\phi$ and $\beta$,
otherwise unachievable by driving the automobiles.

Across all measurements, TX and RX standoff distances to the target were
chosen as short as possible to maximize effective spatial resolution, while
ensuring that each target lay fully within the field-of-view (FoV) of both
arrays. TX/RX heights and standoff distances were selected per target class
to ensure consistent illumination and FoV coverage, with distances for
classes containing multiple targets determined based on the largest target
in the class. The measurement campaign parameters are summarized in
Table~\ref{tab:meascamp}. The remaining target-specific details are listed
below.

\begin{itemize}
\item \textit{UAV:}
A multirotor UAV was mounted on a rotator to vary its
aspect angle. For mechanical stability, the platform was placed on the
ground and inverted on a low stand so that the underside of the UAV could
be observed, which is equivalent to viewing the bottom of the platform in
flight.\footnote{The UAV was inverted solely to allow underside
visibility while ground-mounted; scattering is invariant under this
configuration.} At each TX azimuthal position, the UAV was rotated about
its vertical axis in $5^\circ$ steps from $0^\circ$ to $180^\circ$.
The target reference point used to define TX/RX standoff distances was the
geometric midpoint of the UAV.

\item \textit{Robotic arm:}
An industrial robotic arm with six degrees of freedom was mounted at a height aligned with the TX and RX arrays.
During each acquisition, the arm executed a repeatable L-shaped
trajectory, consisting of a horizontal sweep from lower right to lower
left followed by upward motion. The target reference point used to define
TX/RX standoff distances was the base of the robotic arm.

\item \textit{AMR:}
An autonomous mobile robot was instrumented to
achieve an effective height of 1.49~m, comparable to a humanoid form
factor.\footnote{The platform was outfitted with a rack and equipment
solely to increase its height for experimental purposes; the underlying
mobile base is a standard robotic vehicle.}
The AMR executed two linear 2~m trajectories relative to the TX: a radial
trajectory directed toward the TX and a tangential trajectory
perpendicular to the TX line-of-sight. The target reference point used to
define TX/RX standoff distances was the midpoint of each 2~m trajectory.

\item \textit{Humans:}
Three adult subjects were measured: one female (1.60~m) and two males
(1.69~m and 1.85~m). Each subject executed a straight 2~m walking
trajectory towards the TX, similar to the AMR experiments. The target
reference point used to define TX/RX standoff distances was the midpoint
of the 2~m walking trajectory.

\item \textit{Automobiles:}
Three automobiles were measured: a sedan, a midsize SUV, and a high-roof
van. As explained earlier, the automobiles remained stationary throughout
the measurements and the TX was mobile instead. The target reference point
used to define TX/RX standoff distances was the midpoint of the automobile
body.
\end{itemize}

\section{Measurement System}
\label{sec:sounder}
This section describes the 28~GHz bistatic channel sounder used in this
study. The first subsection describes the RF hardware, the second details the signal processing used to
extract discrete propagation paths from the measurements, and third describes the mapping from the extracted paths to their incident points and associated RCSs for downstream modeling.

\subsection{28~GHz Channel Sounder}
\label{subsec:hardware}

\begin{figure}[!t]
    \centering
    \includegraphics[width=1\linewidth]{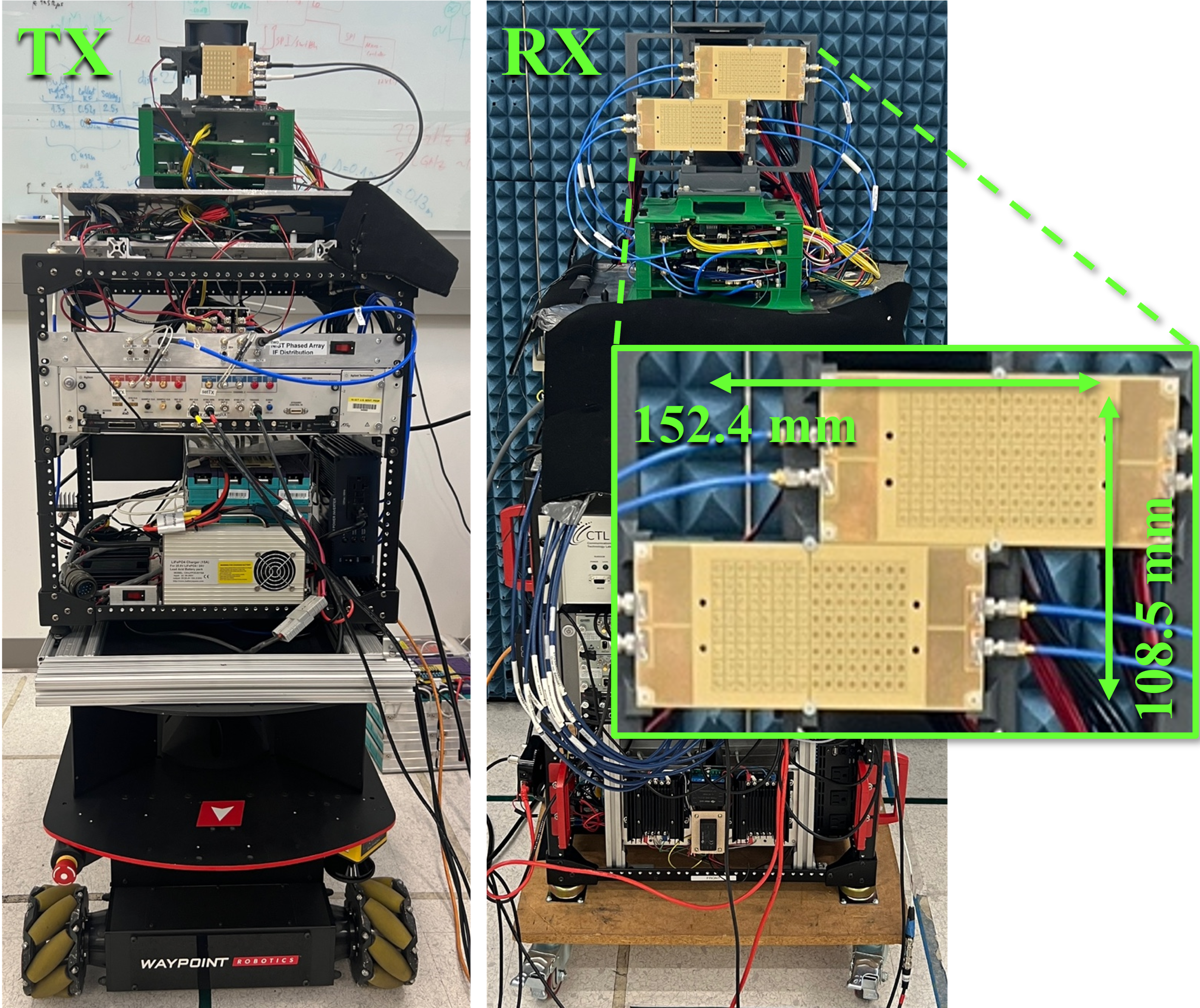}
\caption{Untethered TX and RX of NIST's 28~GHz bistatic channel sounder. The inset shows the full RX aperture formed by four stacked $8\times 8$ phased-array boards, including its overall dimensions.}
    \label{fig:TXRX}
\end{figure}

\begin{figure*}[!t]
\centering

\subfloat[]{
    \includegraphics[width=0.375\textwidth]{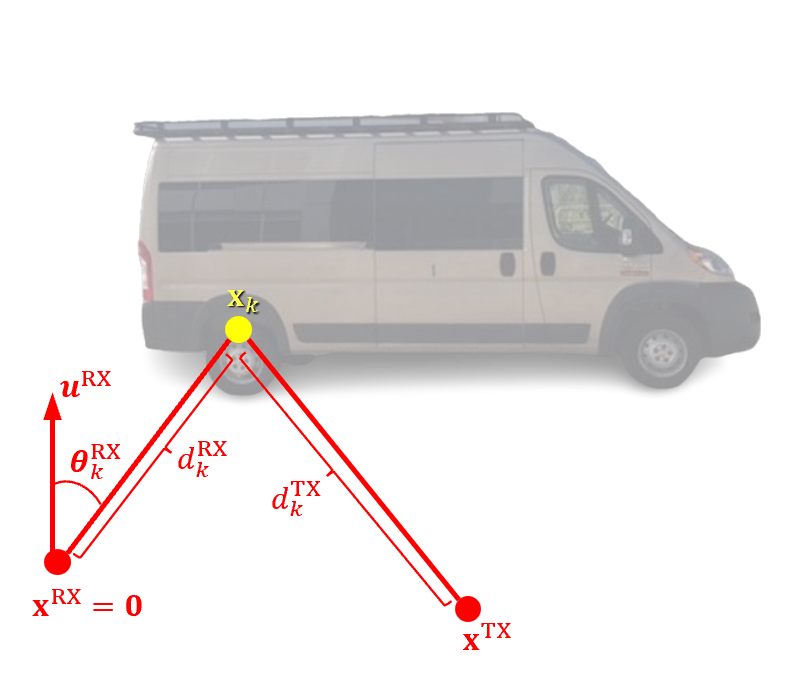}
}
\hfil
\subfloat[]{
    \includegraphics[width=0.375\textwidth]{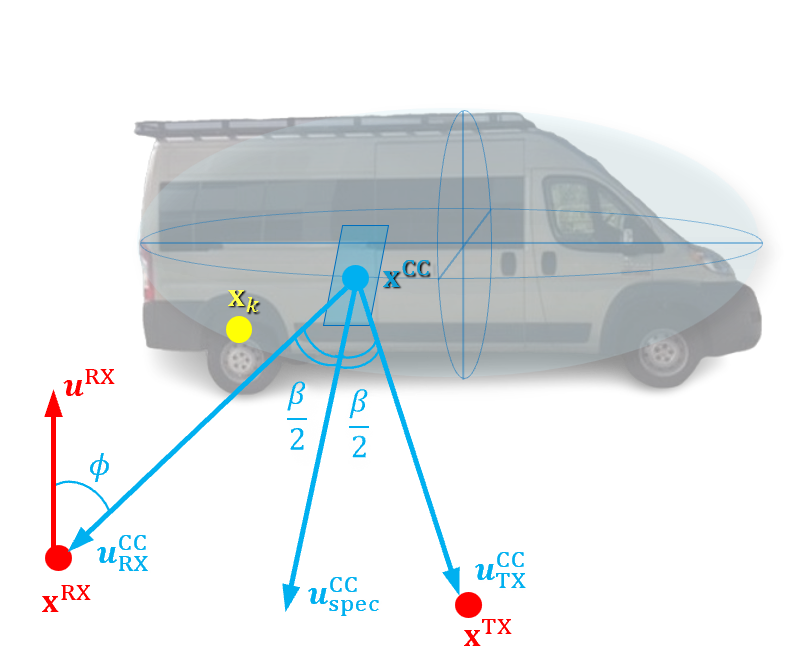}
}
\hfill 
\subfloat[]{
    \includegraphics[width=0.2\textwidth]{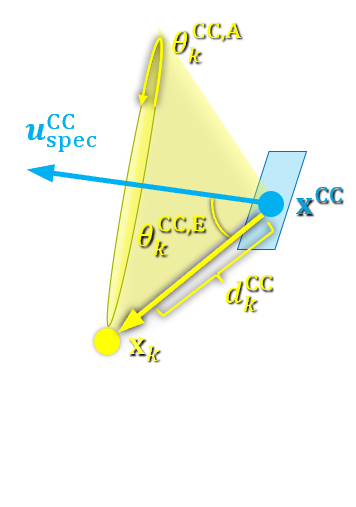}
    
}

\caption{Geometric framework of channel model.
(a) Mapping of a measured path to its incident point on the target
surface.
(b) Cluster center (CC), defined as the intersection of the bistatic specular axis with the canonical ellipsoid representing the target.
(c) Lateral perspective of (b), illustrating the conical distribution of incident points around the bistatic specular axis originating at the cluster center.}
\label{fig:GeometryModel}
\end{figure*}

The measurement system is a 28~GHz channel sounder
(Fig.~\ref{fig:TXRX}) developed at NIST, with complete implementation
details described in~\cite{chuang2024quasi}. The TX and RX operate as
independent platforms, enabling flexible bistatic configurations. 
The TX employs a vertically polarized $8\times 8$ phased-array board
that forms a flat-top beam for target illumination. The RX consists of
four identical $8\times 8$ phased-array boards stacked into a single
256-element aperture for enhanced angular resolution. As shown in the
inset of Fig.~\ref{fig:TXRX}, the full stacked RX aperture spans
152.4~mm by 108.5~mm. After coherent combination of the four boards,
the RX achieves $3.7^\circ$ azimuth and $5.6^\circ$ elevation
beamwidths. Because the boards share a common boresight, the effective
RX aperture retains the same usable FoV as a single board: an azimuth
FoV of $-45^\circ$ to $45^\circ$ and an elevation FoV of $-25^\circ$ to
$25^\circ$.

The channel is probed using a pseudorandom noise (PN) sequence with
2~GHz bandwidth, centered at precisely 28.5~GHz and transmitted at
30~dBm EIRP. The received signal is downconverted to an intermediate
frequency and digitized at 16~GHz. Correlation with the known PN sequence
yields complex-valued channel impulse responses (CIRs) that retain both
amplitude and phase information across all array elements. The CIRs are
calibrated following~\cite{caudill2021real}, achieving delay-domain
sidelobe suppression of at least 55~dB. Timing synchronization for triggering and phase
coherence across acquisitions are maintained via an optical link.

\subsection{Path Extraction}
\label{subsec:path extraction}

The element-level, phase-coherent CIRs obtained at the RX are processed
using the switched-beamforming framework described in
\cite{caudill2021real}. In place of analog beamforming, the CIRs from all
256 RX elements are planar beamformed in post-processing using calibrated
digital weights. The validity of the planar-wave assumption is quantified
in Appendix~\ref{app:near_field}. The calibrated digital weights correct
array non-idealities such as distorted or multi-lobed beams that cannot be
compensated in analog beamforming due to finite-resolution phase shifters.
The digitally synthesized beams achieve angular-domain sidelobe suppression
of at least 18~dB~\cite{caudill2021real}. Because the array elements are
switched rather than the beams, steering to any direction incurs no increase
in channel sweep time, enabling continuous beam synthesis in azimuth and
elevation.

The resulting 3D profile in delay, azimuth AoA, and elevation AoA is
processed by the Space-Alternating Generalized Expectation-Maximization
(SAGE) super-resolution algorithm to extract a discrete set of propagation
paths. Path extraction is therefore performed jointly in delay and 2D angle,
rather than from the angular response alone. The delay-domain sidelobe
suppression of the calibrated CIRs and the angular-domain sidelobe
suppression of the digitally synthesized beams together support the
extraction of weak paths within the measurement-supported dynamic range.
Each path is characterized by a complex amplitude
$\alpha_k = |\alpha_k|e^{j\varphi_k}$, a delay $\tau_k$, and an
angle-of-arrival (AoA) vector
\(
\boldsymbol{\theta}^{\mathrm{RX}}_k
=
(\theta_k^{\mathrm{RX,A}},\,\theta_k^{\mathrm{RX,E}}),
\)
defined with respect to the RX phased-array boresight
$\mathbf{u}^{\mathrm{RX}}$. Together with
$\mathbf{x}^{\mathrm{RX}}=\mathbf{0}$, this unit vector defines the global
coordinate system illustrated in Fig.~\ref{fig:GeometryModel}. Antenna and
RF-chain responses are de-embedded following~\cite{caudill2021real}, such
that $\alpha_k$ reflects propagation effects only. Although the inherent RX
beamwidths are $3.7^\circ$ (azimuth) and $5.6^\circ$ (elevation), SAGE
improves both delay and angle resolution by more than a factor of five
\cite{papazian2016calibration, chuang2024quasi}. In the controlled
validation experiment summarized in Appendix~\ref{app:path_validation},
this extraction pipeline yielded average estimation errors of $0.1$~ns in
delay and $0.2^\circ$ in AoA.

\subsection{Path-to--Incident-Point Mapping}
\label{subsec:path_rcs}
In their native form, the measured path properties—complex amplitude
$|\alpha_k|e^{j\varphi_k}$, delay $\tau_k$, and AoA
$\boldsymbol{\theta}^{\mathrm{RX}}_k$—are not directly interpretable in
geometric terms. To ensure physically grounded spatio-temporal
consistency, the proposed channel model is formulated in terms of
incident points rather than paths. Each extracted path is therefore
mapped to a point $\mathbf{x}_k$ on the target surface and to a
scattering strength expressed by an intrinsic bistatic RCS $\sigma_k$.
Together with the measured phase $\varphi_k$, these quantities define
the scattering primitives used in the subsequent sections.

The mapping is implemented using the bistatic range
surface--line construction~\cite{10999612}, illustrated in
Fig.~\ref{fig:GeometryModel}(a). For each path $k$, the TX, RX, and
incident point $\mathbf{x}_k$ form a bistatic triangle where
the measured delay $\tau_k$ constrains the total TX--incident point--RX path
length through the bistatic range condition
\begin{equation}
d_k^{\mathrm{TX}} + d_k^{\mathrm{RX}} = c\,\tau_k ,
\label{eq:bistatic_range_constraint}
\end{equation}
where $c$ is the speed of light, and
\begin{subequations}\label{eq:tx_rx_distance_def}
\begin{align}
d_k^{\mathrm{TX}} &= \|\mathbf{x}_k-\mathbf{x}^{\mathrm{TX}}\|
\label{eq:tx_distance_def} \\
d_k^{\mathrm{RX}} &= \|\mathbf{x}_k-\mathbf{x}^{\mathrm{RX}}\|.
\label{eq:rx_distance_def}
\end{align}
\end{subequations}
Since $\mathbf{x}^{\mathrm{RX}}=\mathbf{0}$, the AoA
$\boldsymbol{\theta}^{\mathrm{RX}}_k$ directly defines the RX-side ray,
and the incident point lies along this ray as
\begin{equation}
\mathbf{x}_k
=
d_k^{\mathrm{RX}}
\begin{bmatrix}
\cos\theta_k^{\mathrm{RX,A}}\,\cos\theta_k^{\mathrm{RX,E}} \\
\sin\theta_k^{\mathrm{RX,A}}\,\cos\theta_k^{\mathrm{RX,E}} \\
\sin\theta_k^{\mathrm{RX,E}}
\end{bmatrix}.
\label{eq:path_cartesian_location}
\end{equation}
Substituting \eqref{eq:path_cartesian_location} into
\eqref{eq:tx_distance_def} and enforcing
\eqref{eq:bistatic_range_constraint} yields a scalar equation in
$d_k^{\mathrm{RX}}$, whose solution determines the incident point
$\mathbf{x}_k$.

Complete derivation and solution details for the mapping are provided
in~\cite{10999612}. The bistatic range surface--line construction is
strictly valid for first-order interactions, which is the case here
since the proposed model focuses on first-order scatter from the
target. To retain only paths whose mapped points lie within
the target bounds, 3D gating is applied. The target was intentionally positioned with
sufficient spatial separation from other objects to enable this
geometric discrimination. Higher-order paths from the target
surroundings are not physically interpreted as true incident points, but
they nevertheless map outside the known target volume and are therefore
filtered out by the gating.

The incident-point RCS is obtained via the calibrated bistatic radar
equation. Absolute scaling is established through the metal-sphere
calibration~\cite{10999612}. After calibration, the effective TX power
and antenna gains satisfy
$P^{\mathrm{TX}} = G^{\mathrm{TX}} = G^{\mathrm{RX}} = 1$
in linear scale. Under these conditions, the bistatic radar equation
reduces to
\begin{equation}
|\alpha_k|^2
=
\frac{\lambda^2\,\sigma_k}
{(4\pi)^3\,(d_k^{\mathrm{TX}})^2\,(d_k^{\mathrm{RX}})^2},
\label{eq:path_radar_equation}
\end{equation}
where $\lambda=c/f_c$ denotes the wavelength corresponding to the
center frequency $f_c$. Solving for the incident-point RCS yields
\begin{equation}
\sigma_k
=
\frac{(4\pi)^3 (d_k^{\mathrm{TX}})^2 (d_k^{\mathrm{RX}})^2}{\lambda^2}
\,|\alpha_k|^2 .
\label{eq:path_rcs}
\end{equation}
The phase $\varphi_k$ associated with each path is retained to enable
coherent superposition in the computation of the \emph{composite RCS} in
Section~\ref{subsec:composite_rcs}, which represents the total
scattering strength.

\begin{figure}[!t]
  \centering
  \includegraphics[width=\linewidth]{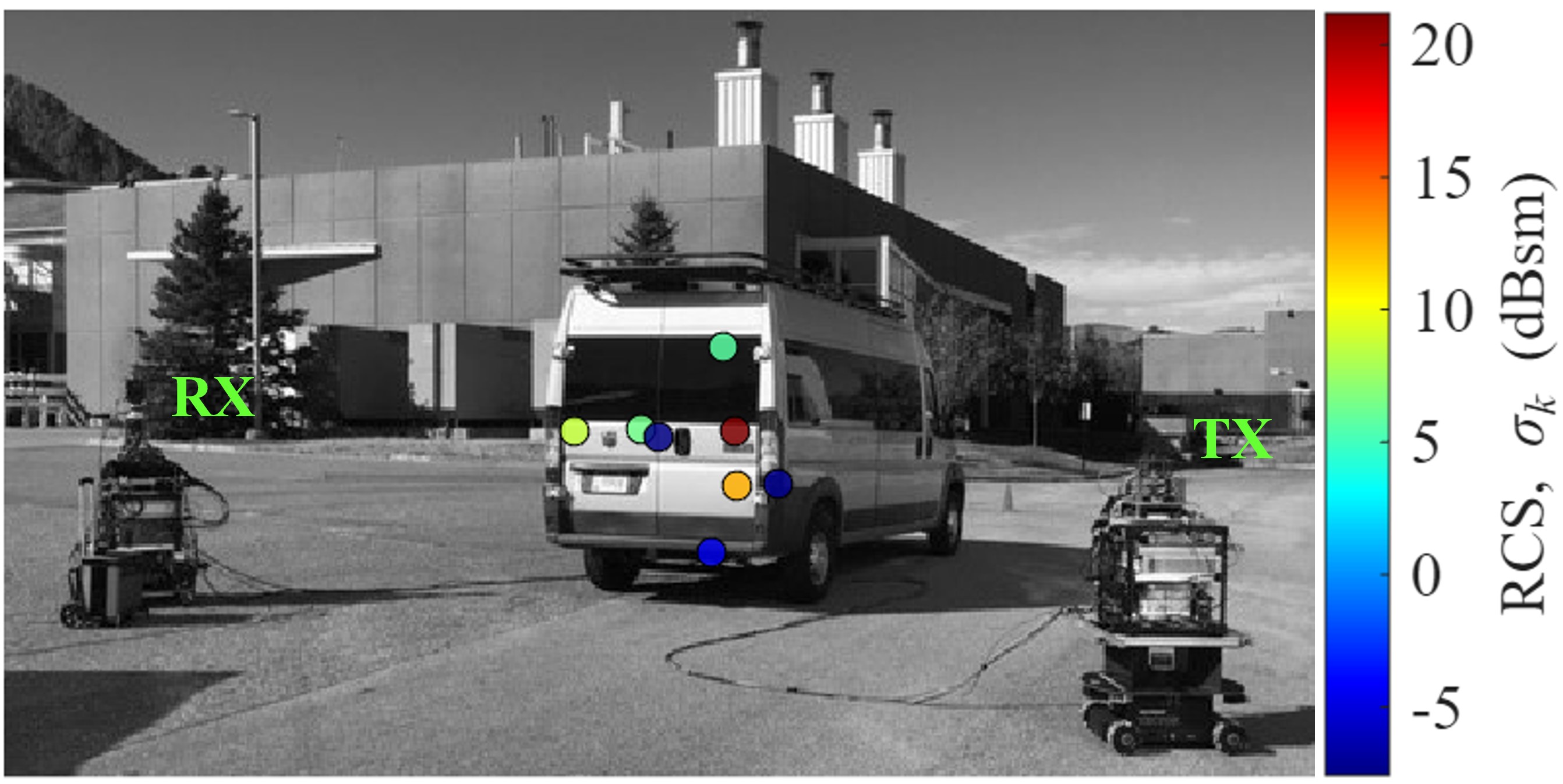}
  \caption{Visualization of eight detected paths for a single acquisition of the van target, mapped to incident points on its surface. Marker color denotes the estimated radar cross section (RCS).}
  \label{fig:van_mpcs}
\end{figure}

Fig.~\ref{fig:van_mpcs} shows a representative snapshot of the eight paths
extracted for a single acquisition of the van target, mapped to incident
points and projected onto the corresponding camera image via Lidar
thanks to our context-aware channel sounder~\cite{gentile2024context}. The
spatial dispersion of the points across the van highlights the
distributed nature of the scattering mechanisms, with contributions
arising from multiple structural features such as the rear door, bumper,
windows, and underbody. The strongest path within this set corresponds
to the vicinity of the bistatic-specular scattering location and appears
as the dominant member of the cluster.

\section{Channel Model: Geometry}
\label{sec:cm_geom}
This section describes the geometric component of the proposed model, in
which target scattering is represented as a cluster of paths anchored
at the specular reflection point on the target. This point is introduced
in the first subsection as the \emph{cluster center} (CC), a deterministic
bistatic--specular reflection point defined by the TX--target--RX geometry.
The cluster shape is then characterized by three stochastic parameters:
path count, distance offset, and angle offset. These parameters are
introduced in the following subsections, and the resulting model
parameters are analyzed in the final subsection.

\subsection{Cluster Center}
\label{subsec:sc_definition}

Measurements show that, for each acquisition, the strongest path most
often originates from the surface location satisfying the bistatic
specular condition for the given
$\mathrm{TX}$--target--$\mathrm{RX}$ geometry. For convex targets at
mmWave frequencies, this specular contribution dominates and serves as a
natural anchor, while nearby diffuse scattering forms a compact cluster
around it \cite{gentile2018quasi,varshney2021quasi}. As illustrated in
Fig.~\ref{fig:GeometryModel}(b), this location is associated with the
surface normal that bisects the incident and reflected rays.

To define this location uniformly across heterogeneous targets, each
target is represented by a canonical ellipsoid whose principal axes
match its measured height, length, and depth. This ellipsoid is used as a
reduced-order geometric reference rather than as a detailed surface model
of the target. Although it inevitably simplifies non-convex, articulated,
or highly irregular targets, this choice is consistent with the objective
of a geometric--stochastic channel model, which seeks a compact
parametric representation rather than a high-fidelity raytracing or
electromagnetic-solver solution. Deviations between the actual target shape and
the canonical ellipsoid are captured by the stochastic component of the
model defined in the following subsections.

For a given TX and RX
placement, the intersection of the bistatic specular axis with the
ellipsoid defines a unique surface point
$\mathbf{x}^{\mathrm{CC}}$, which is referred to as the cluster center
(CC).
Next, let
\(
\mathbf{u}^{\mathrm{CC}}_{\mathrm{TX}}
=
\mathbf{x}^{\mathrm{TX}}-\mathbf{x}^{\mathrm{CC}}
\)
and
\(
\mathbf{u}^{\mathrm{CC}}_{\mathrm{RX}}
=
\mathbf{x}^{\mathrm{RX}}-\mathbf{x}^{\mathrm{CC}}
\)
denote the vectors from the cluster center toward the TX and RX, respectively. The
bistatic specular axis is defined as the bisector of these two vectors,
\begin{equation}
\mathbf{u}^{\mathrm{CC}}_{\mathrm{spec}}
=
\frac{\mathbf{u}^{\mathrm{CC}}_{\mathrm{TX}}}
{\|\mathbf{u}^{\mathrm{CC}}_{\mathrm{TX}}\|}
+
\frac{\mathbf{u}^{\mathrm{CC}}_{\mathrm{RX}}}
{\|\mathbf{u}^{\mathrm{CC}}_{\mathrm{RX}}\|}.
\label{eq:specular_axis}
\end{equation}
This axis acts as the geometric backbone of the cluster,
as depicted in Fig.~\ref{fig:GeometryModel}(b).

\subsection{Path Count}
\label{subsec:path count}
The number of scattered paths around the cluster center varies stochastically
across acquisitions. For each target, the per-acquisition path counts
across all measurements were aggregated into an empirical distribution and
fit with the Poisson model. This conclusion is supported by
Kolmogorov--Smirnov (KS) goodness-of-fit testing across all targets:
\begin{equation}
    f(N) = \frac{\lambda^{N} e^{-\lambda}}{N!},
    \label{eq:distribution path count}
\end{equation}
where the mean $\lambda$ is target dependent
(Table~\ref{tab:geom_params}). Representative empirical distributions and
Poisson fits appear in Fig.~\ref{fig:PathCount}.

The same aggregation and fitting procedure, together with the KS test
against candidate distributions, is used for the distance offset described
next and for all subsequent statistical characterizations in this paper.

\subsection{Distance Offset}
\label{subsec:distance offset}

For each path $k$, let the distance offset 

\begin{equation}
d_k^{\mathrm{CC}}
=
\bigl\|\mathbf{x}_k - \mathbf{x}^{\mathrm{CC}}\bigr\|
\label{eq:distance_offset_def}
\end{equation}
denote the radial deviation of incident point $\mathbf{x}_k$ from the cluster center $\mathbf{x}^\mathrm{CC}$,
as illustrated in Fig.~\ref{fig:GeometryModel}(c).
Across all targets, $d_k^{\mathrm{CC}}$ was found to obey the Gamma distribution
\begin{equation}
f(d^\mathrm{CC})
=
\frac{(d^\mathrm{CC})^{\,a-1} e^{-d^\mathrm{CC}/b}}{b^{a}\,\Gamma(a)},
\label{eq:dist_dsc}
\end{equation}
with target-dependent parameters $(a,b)$ listed in
Table~\ref{tab:geom_params}. Representative fits are shown in
Fig.~\ref{fig:Distance offset}.

%This distance offset constitutes one of the two intrinsic coordinates
%that characterize the local scattering cluster. The second coordinate,
%which captures angular deviation relative to the specular axis, is described next.

\subsection{Angle Offset}
\label{subsec:angle offset}
For each path $k$, let
$\mathbf{u}^{\mathrm{CC}}_k=\mathbf{x}_k-\mathbf{x}^{\mathrm{CC}}$
denote the vector from the cluster center toward its reconstructed
incident point. The orientation of this vector relative to the bistatic
specular axis $\mathbf{u}^{\mathrm{CC}}_{\mathrm{spec}}$ is parameterized
by the 2D angle offset
\begin{equation}
\boldsymbol{\theta}^{\mathrm{CC}}_k
=
\bigl(\theta_k^{\mathrm{CC,A}},\,\theta_k^{\mathrm{CC,E}}\bigr).
\label{eq:angle_offset_def}
\end{equation}
The elevation-like deviation from the specular axis is defined as
\begin{equation}
\theta_k^{\mathrm{CC,E}}
=
\cos^{-1}\!\left(
\frac{
\mathbf{u}^{\mathrm{CC}}_{\mathrm{spec}}
\cdot
\mathbf{u}^{\mathrm{CC}}_k
}{
\|\mathbf{u}^{\mathrm{CC}}_{\mathrm{spec}}\|\,
\|\mathbf{u}^{\mathrm{CC}}_k\|
}
\right),
\label{eq:elevation_angle_offset}
\end{equation}
where $\theta_k^{\mathrm{CC,E}}=0^\circ$ corresponds to perfect alignment
with the specular axis and larger values indicate increasing angular
deviation.

Across all targets, the empirical distribution of
$\theta_k^{\mathrm{CC,E}}$ is strongly forward-weighted, with maximum
density near $0^\circ$ and monotonic decay with increasing angle. This
behavior is well modeled by
\begin{equation}
f(\theta^{\mathrm{CC,E}})
=
\lambda\,\exp\!\left(-\lambda\,\theta^{\mathrm{CC,E}}\right),
\label{eq:dist_ang_offset}
\end{equation}
with decay rates summarized in Table~\ref{tab:geom_params} and illustrated
in Fig.~\ref{fig:AngleOffset}.
The azimuthal component $\theta_k^{CC,A}$ specifies rotation about the specular axis and is approximately uniform over $[0^\circ,360^\circ)$, as shown empirically in Fig.~\ref{fig:AzOffset}.

Together with the distance offset $d_k^{\mathrm{CC}}$, the elevation
offset $\theta_k^{\mathrm{CC,E}}$ defines a conical surface centered on
the specular axis. The azimuth offset then selects a unique incident
point on this surface. This geometry is depicted in
Fig.~\ref{fig:GeometryModel}(c), which provides a lateral perspective of
the configuration shown in Fig.~\ref{fig:GeometryModel}(b).

\begin{table*}[h!]
\centering
\caption{Geometry model parameters.}
\label{tab:geom_params}
\renewcommand{\arraystretch}{1.3}
\setlength{\tabcolsep}{6pt}

\begin{tabular}{l|c|c|c}
\hline
Target &
\shortstack{Path Count, $N$\\$\lambda$ (see (\ref{eq:distribution path count}))} &
\shortstack{Distance Offset, $d_k^{\mathrm{CC}}$\\$(a,b)$ (see (\ref{eq:dist_dsc}))} &
\shortstack{Elevation Angle Offset, $\theta^{\mathrm{CC,E}}$\\$\lambda$ (see (\ref{eq:dist_ang_offset}))} \\
\hline\hline
UAV                & 1.54 & 1.89, 0.10 & 0.026 \\
Robotic Arm        & 2.67 & 2.38, 0.07 & 0.033 \\
AMR                & 4.43 & 2.61, 0.13 & 0.029 \\
Human--1.60 m F    & 2.40 & 0.90, 0.35 & 0.043 \\
Human--1.69 m M    & 4.38 & 1.44, 0.23 & 0.052 \\
Human--1.85 m M    & 4.92 & 1.86, 0.17 & 0.055 \\
Automobile--Sedan  & 3.12 & 0.94, 0.68 & 0.035 \\
Automobile--SUV    & 5.06 & 1.54, 0.50 & 0.035 \\
Automobile--Van    & 6.64 & 1.83, 0.73 & 0.033 \\
\bottomrule
\end{tabular}

%\vspace{0.9em}

% ---------- FIGURE 1: PATH COUNT ----------
\begin{minipage}{\textwidth}
    \centering
    \setcounter{subtable}{0}

    \subfloat[UAV]{%
        \includegraphics[width=0.2\textwidth]{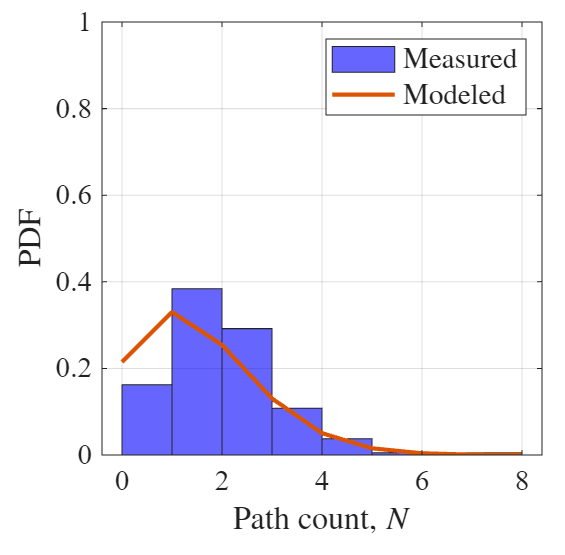}%
        \label{fig:PathCount_UAV}%
    }
    \hfill
    \subfloat[Robotic Arm]{%
        \includegraphics[width=0.2\textwidth]{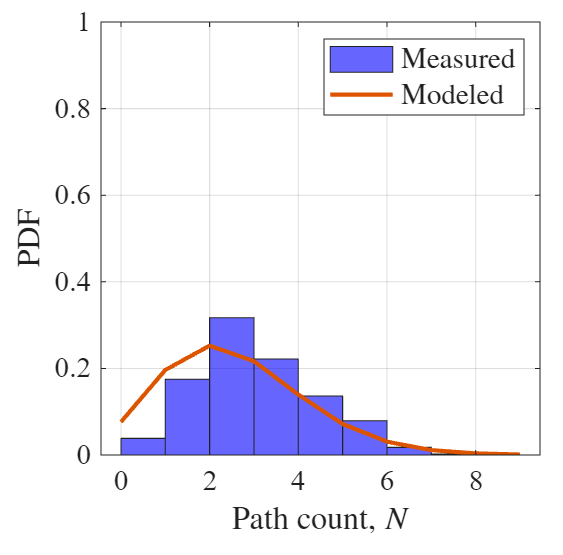}%
        \label{fig:PathCount_Arm}%
    }
    \hfill
    \subfloat[AMR]{%
        \includegraphics[width=0.2\textwidth]{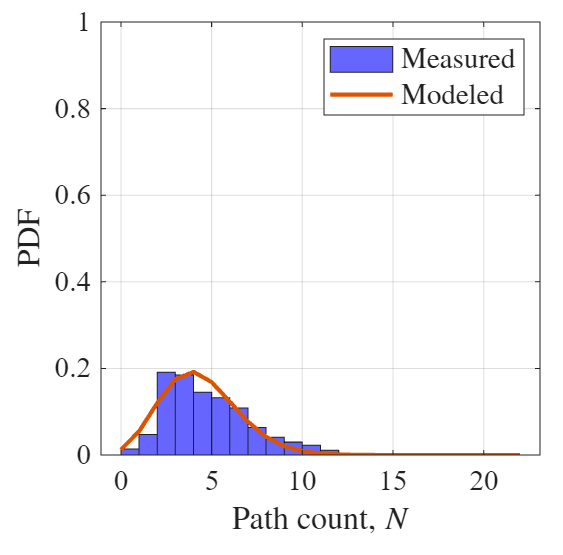}%
        \label{fig:PathCount_AMR}%
    }
    \hfill
    \subfloat[Human--1.85\,m Male]{%
        \includegraphics[width=0.2\textwidth]{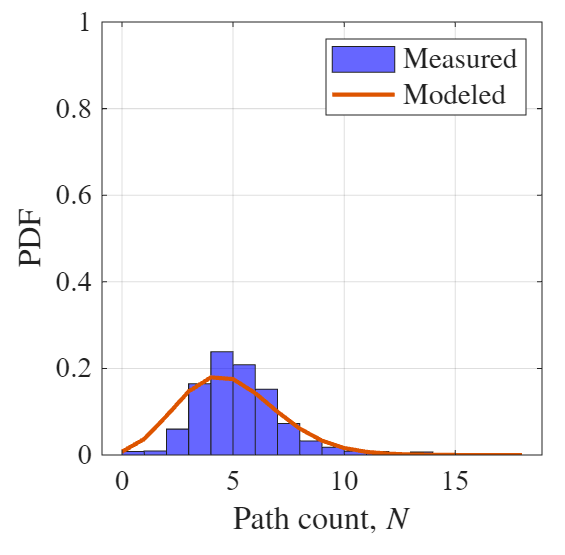}%
        \label{fig:PathCount_Jihoon}%
    }
    \hfill
    \subfloat[Automobile--Van]{%
        \includegraphics[width=0.2\textwidth]{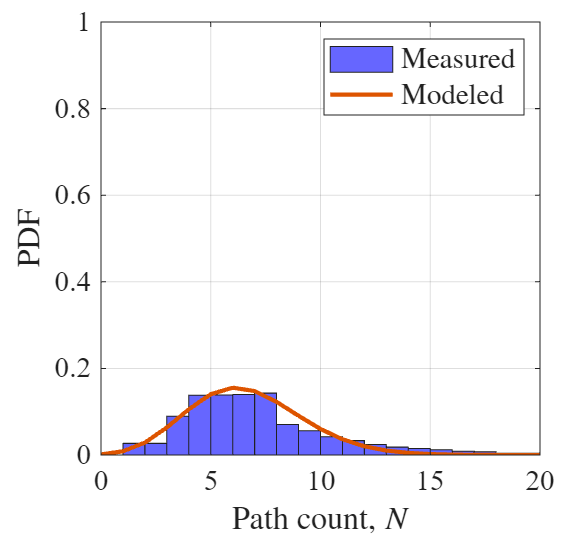}%
        \label{fig:PathCount_Van}%
    }

    \refstepcounter{figure}
    %\vspace{0.5em}

    \raggedright
    {Fig.~\thefigure. Geometry model: Empirical and fitted distributions of the path count $N$ for five representative targets.}
    \label{fig:PathCount}
\end{minipage}

%\vspace{0.9em}

% ---------- FIGURE 2: DISTANCE OFFSET ----------
\begin{minipage}{\textwidth}
    \centering
    \setcounter{subtable}{0}

    \subfloat[UAV]{%
        \includegraphics[width=0.19\textwidth]{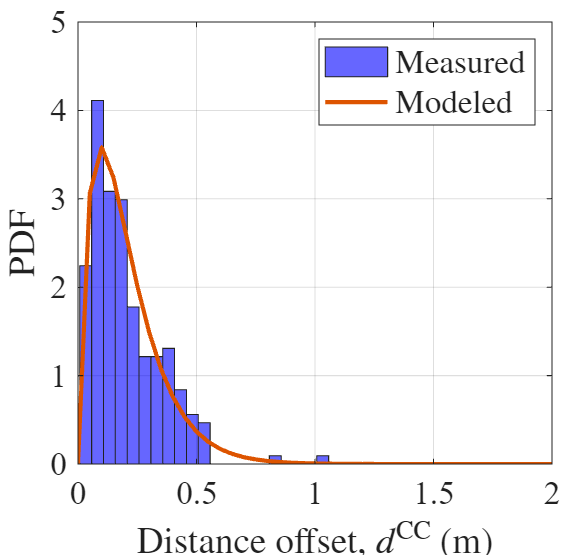}%
        \label{fig:RadialOffset_UAV}%
    }
    \hfill
    \subfloat[Robotic Arm]{%
        \includegraphics[width=0.19\textwidth]{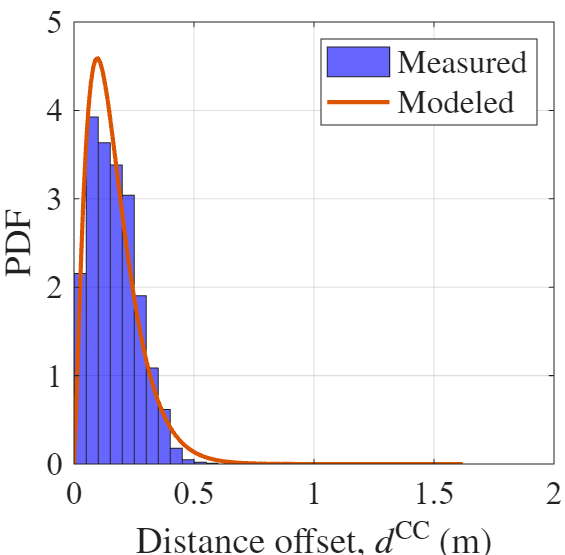}%
        \label{fig:RadialOffset_Arm}%
    }
    \hfill
    \subfloat[AMR]{%
        \includegraphics[width=0.19\textwidth]{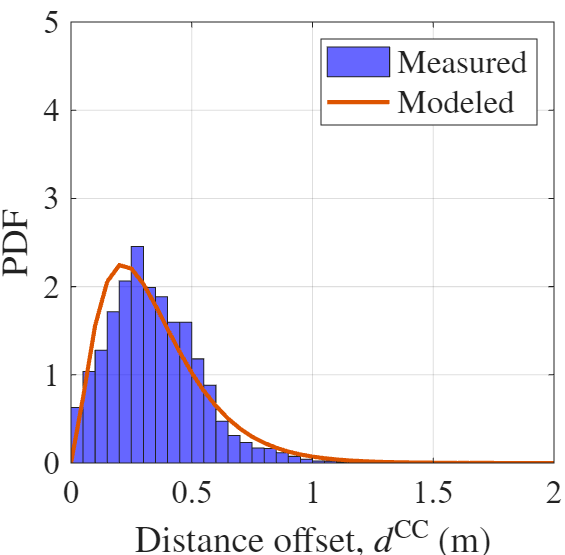}%
        \label{fig:RadialOffset_AMR}%
    }
    \hfill
    \subfloat[Human--1.85\,m Male]{%
        \includegraphics[width=0.19\textwidth]{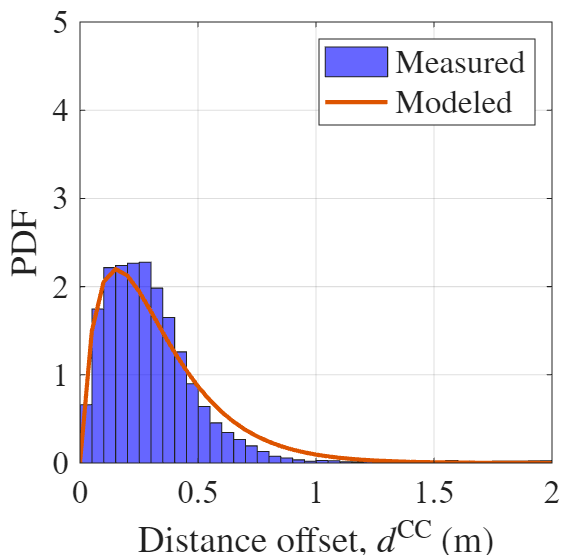}%
        \label{fig:RadialOffset_Jihoon}%
    }
    \hfill
    \subfloat[Automobile--Van]{%
        \includegraphics[width=0.19\textwidth]{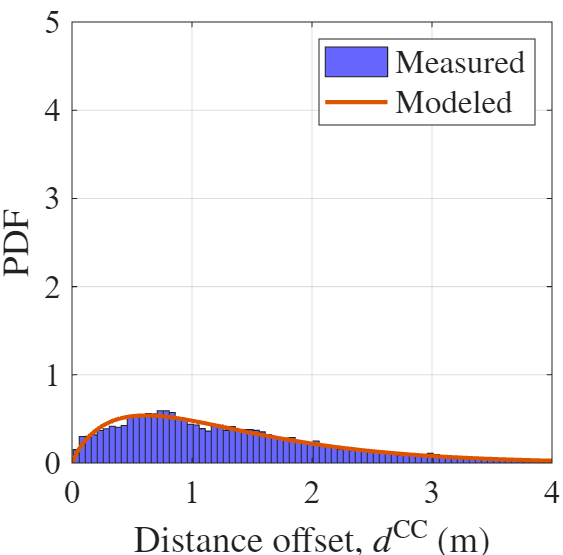}%
        \label{fig:RadialOffset_Van}%
    }

    \refstepcounter{figure}
    %\vspace{0.5em}

    \raggedright
    {Fig.~\thefigure. Geometry model: Empirical and fitted distributions of the distance offset $d_k^{\mathrm{CC}}$ for five representative targets.}
    \label{fig:Distance offset}
\end{minipage}

%\vspace{0.9em}

% ---------- FIGURE 3: ANGLE OFFSET ----------
\begin{minipage}{\textwidth}
    \centering
    \setcounter{subtable}{0}

    \subfloat[UAV]{%
        \includegraphics[width=0.2\textwidth]{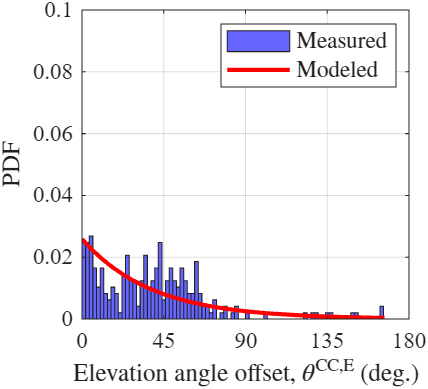}%
        \label{fig:AngleOffset_UAV}%
    }
    \hfill
    \subfloat[Robotic Arm]{%
        \includegraphics[width=0.2\textwidth]{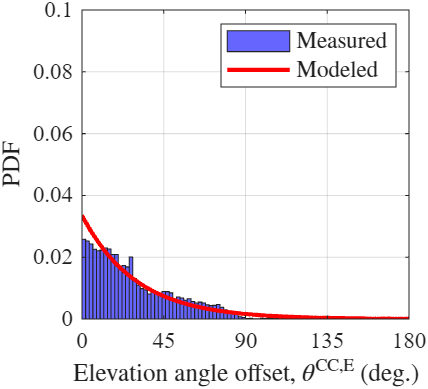}%
        \label{fig:AngleOffset_Arm}%
    }
    \hfill
    \subfloat[AMR]{%
        \includegraphics[width=0.2\textwidth]{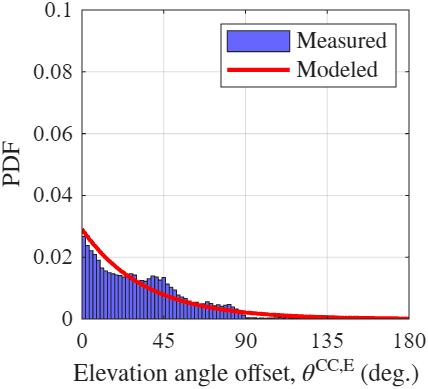}%
        \label{fig:AngleOffset_AMR}%
    }
    \hfill
    \subfloat[Human--1.85\,m Male]{%
        \includegraphics[width=0.2\textwidth]{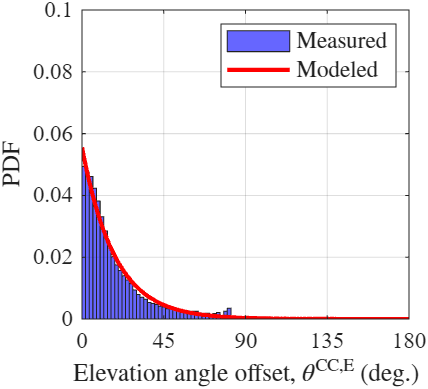}%
        \label{fig:AngleOffset_Jihoon}%
    }
    \hfill
    \subfloat[Automobile--Van]{%
        \includegraphics[width=0.2\textwidth]{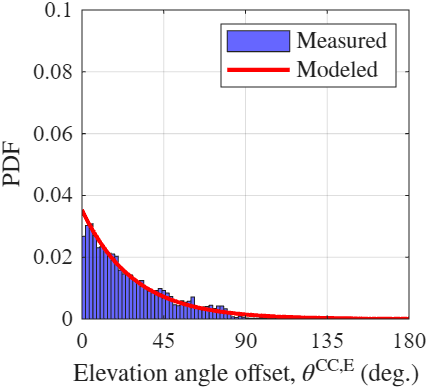}%
        \label{fig:AngleOffset_Van}%
    }

    \refstepcounter{figure}
    %\vspace{0.5em}

    \raggedright
    {Fig.~\thefigure. Geometry model: Empirical and fitted distributions of the elevation angle offset $\theta_k^{\mathrm{CC,E}}$ for five representative targets.}
    \label{fig:AngleOffset}
\end{minipage}

% ---------- FIGURE 8: AZ ANGLE OFFSET ----------
\begin{minipage}{\textwidth}
    \centering
    \setcounter{subtable}{0}

    \subfloat[UAV]{%
        \includegraphics[width=0.2\textwidth]{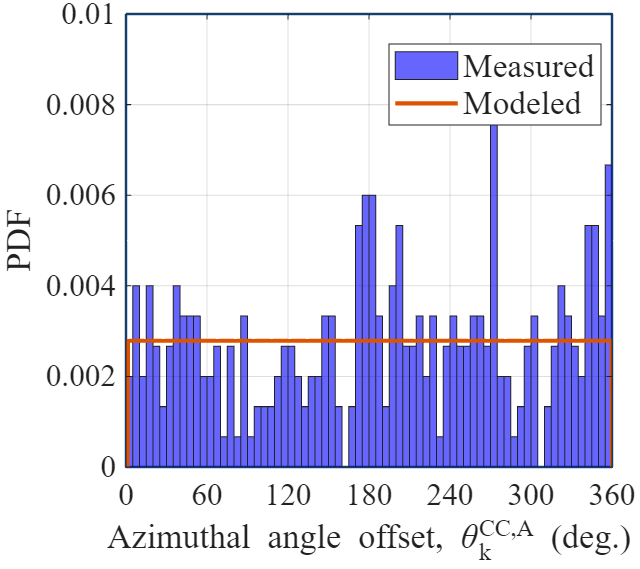}%
        \label{fig:AzOffset_UAV}%
    }
    \hfill
    \subfloat[Robotic Arm]{%
        \includegraphics[width=0.2\textwidth]{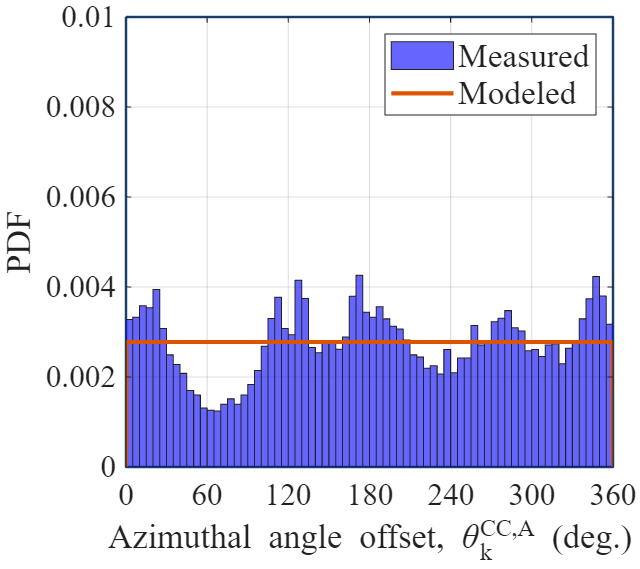}%
        \label{fig:AzOffset_Arm}%
    }
    \hfill
    \subfloat[AMR]{%
        \includegraphics[width=0.2\textwidth]{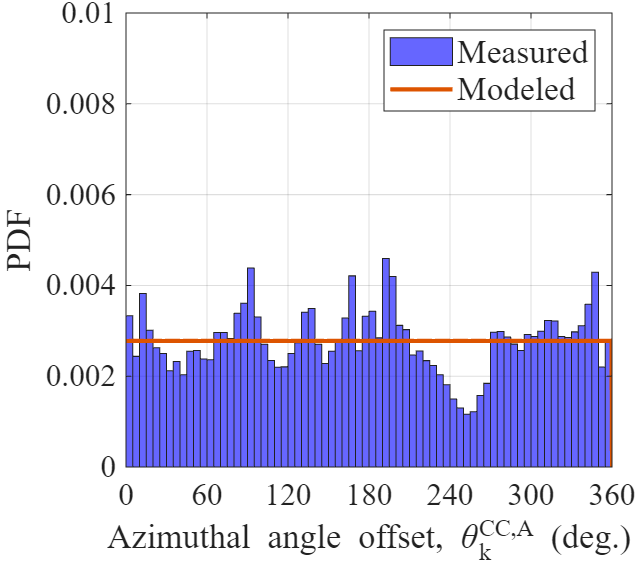}%
        \label{fig:AzOffset_AMR}%
    }
    \hfill
    \subfloat[Human--1.85\,m Male]{%
        \includegraphics[width=0.2\textwidth]{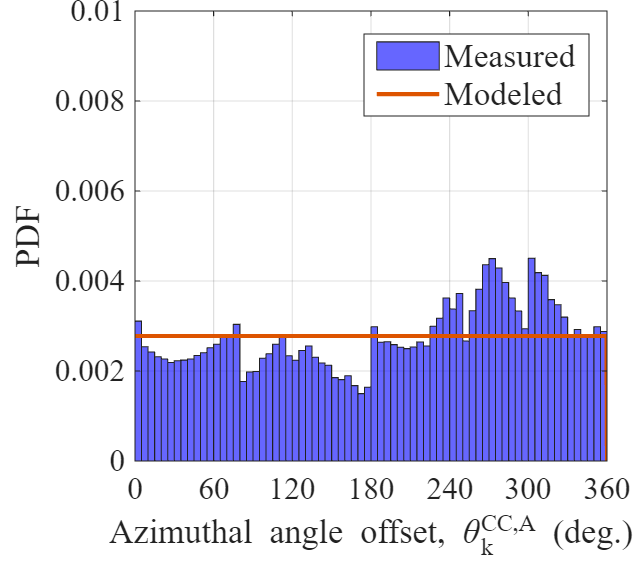}%
        \label{fig:AzOffset_Jihoon}%
    }
    \hfill
    \subfloat[Automobile--Van]{%
        \includegraphics[width=0.2\textwidth]{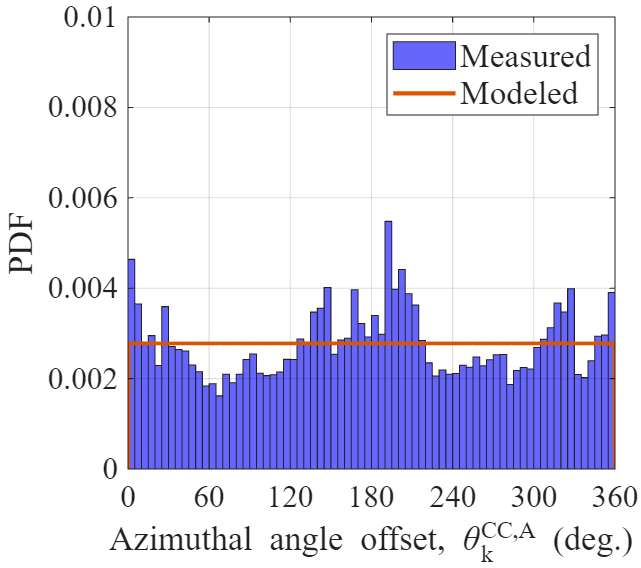}%
        \label{fig:AzOffset_Van}%
    }

    \refstepcounter{figure}
    %\vspace{0.5em}

    \raggedright
    {Fig.~\thefigure. Geometry model: Empirical and fitted distributions of the azimuthal angle offset $\theta_k^{\mathrm{CC,A}}$ for five representative targets.}
    \label{fig:AzOffset}
\end{minipage}

\end{table*}

\subsection{Parameter Analysis}
\label{subsec:joint_geometry}

Across the nine measured targets, the fitted 
parameters in Table~\ref{tab:geom_params} and the representative 
distributions in Figs.~\ref{fig:PathCount}--\ref{fig:AngleOffset} reveal 
clear cross-target patterns driven by size, curvature, surface 
composition, and geometric complexity.

\vspace{0.3em}\noindent
\textit{Path count:}
As established earlier, the number of resolvable paths per
acquisition follows a Poisson distribution characterized by a
target-dependent rate parameter $\lambda$. 
For both humans and automobiles, $\lambda$ increases monotonically with
target size: the largest human and largest automobile produce the most
paths, medium-sized subjects fall in the middle, and the smallest yield
the fewest. The UAV and robotic arm generate the lowest path counts
overall, consistent with their compact dimensions. The AMR falls between
these extremes, reflecting its intermediate size and surface extent.
A notable exception is the robotic arm, which produces more paths than
its physical size alone would suggest. Its metallic, articulated, and
multi-faceted structure introduces additional sub-specular scattering
mechanisms that elevate $\lambda$ beyond what would be expected from
geometry alone.

\vspace{0.3em}\noindent
\textit{Distance offset:} The Gamma-distributed distance offsets
$d_k^{\mathrm{CC}}$ reflect how far resolvable paths extend around the cluster center. 
The fitted scale parameters follow intuitive
geometric patterns: larger targets—particularly the van and SUV, and to a
lesser degree the AMR—produce the widest spreads because their extended
surfaces support sub-specular returns over a broader region. Human
subjects again follow a monotonic size ordering—tallest to shortest—
matching the visible torso widths in Figs.~\ref{fig:targets9}(d-f). Compact
platforms such as the UAV and robotic arm generate the tightest
distributions, consistent with highly localized scattering zones. As with
the path-count behavior, the robotic arm shows slightly larger spread
than expected for its size due to its complex metallic geometry, which
generates multiple nearby reflections around the specular point. These
observations demonstrate that the Gamma parameters effectively encode how
physical dimensions and surface structure determine the radial width of
the scattering cluster.

\vspace{0.3em}\noindent
\textit{Angle offset:}
The angle-offset statistics reveal a physically consistent organization
of the scattering cluster across all targets. As established in
Section~\ref{subsec:angle offset}, azimuthal offsets
$\theta_k^{\mathrm{CC,A}}$ are approximately uniform about the specular axis, while
elevation offsets $\theta_k^{\mathrm{CC,E}}$ are strongly
forward-weighted, with maximum density at zero degrees. The emergence of
these simple, well-behaved distributions would not occur if the
deterministic geometry were misaligned with the underlying scattering
mechanisms. Their consistency across disparate target classes therefore
confirms that the cluster-centered geometry captures the dominant scattering structure without introducing artificial anisotropy.

The fitted exponential decay rates $\lambda$ governing
$\theta_k^{\mathrm{CC,E}}$ further quantify the angular concentration of
incident points around the specular axis. Since smaller $\lambda$ corresponds
to slower decay and therefore broader angular spread, the automobiles
exhibit relatively broad angular spreads, consistent with their large
illuminated surfaces; however, their comparatively smooth body shapes
suggest that this spread is driven primarily by physical extent rather than
by geometric complexity. By contrast, the human subjects exhibit the largest
decay rates and therefore the most concentrated angular spreads, consistent
with their smaller size, non-metallic composition, and comparatively smooth
body geometry. The robotic targets exhibit broad angular spreads despite
their smaller dimensions, which can be attributed to their metallic,
articulated, and geometrically complex structures that introduce multiple
off-specular scattering contributions. Thus, the angle-offset parameter
captures a combined effect of size, material composition, and geometric
complexity, rather than target size alone.

\section{Channel Model: RCS}
\label{sec:cm_rcs}
This section describes the RCS component of the
proposed model, which assigns scattering strength to the incident-point
geometry described in the previous section. It was formulated
to remain compatible with 3GPP TR~38.901, the de facto channel model used
for wireless system simulation and design. This compatibility allows it to be
incorporated directly into 3GPP-style ISAC channel simulations and
system-level performance evaluations.
Accordingly, the RCS is first modeled at the cluster level through the
{\em composite} RCS, which represents the total scattered strength
across all incident points and follows the 3GPP-style decomposition into
mean, deterministic geometry-dependent, and stochastic fluctuation terms.
The second subsection introduces a redistribution model that maps the
composite RCS to individual incident-point RCS values, thereby coupling the
cluster-level RCS characterization to the path-specific geometry. The final
subsection interprets the fitted RCS parameters across targets.

\subsection{Composite RCS}
\label{subsec:composite_rcs}
For each acquisition, the \emph{composite RCS} represents the total
coherent scattering strength of the incident points on the target. Given the $N$ paths along with their
corresponding phases $\varphi_k$ and incident-point RCS values $\sigma_k$
obtained from (\ref{eq:path_rcs}), the composite RCS is computed as
\begin{equation}
    \sigma
    =
    \left|
        \sum_{k=1}^{N}
        \sqrt{\sigma_k}\, e^{j(\varphi_k+2\pi f_c\tau_k)}
    \right|^2 ,
    \label{eq:composite_rcs_new}
\end{equation}
where all RCS quantities are expressed in linear units (m$^2$) unless
stated otherwise.
Following 3GPP~TR~38.901, the composite RCS is modeled using a
multiplicative decomposition as a function of bistatic angle $\beta$ and
aspect angle $\phi$,
\begin{equation}
    \sigma(\beta,\phi)
    =
    \sigma_{\mathrm{M}}\,
    \sigma_{\mathrm{D}}(\beta,\phi)\,
    \sigma_{\mathrm{S}},
    \label{eq:rcs_mult_new}
\end{equation}
where $\sigma_{\mathrm{M}}$ denotes the {\em mean} composite RCS,
$\sigma_{\mathrm{D}}(\beta,\phi)$ the {\em deterministic} composite RCS capturing angular
dependence, and $\sigma_{\mathrm{S}}$ the {\em stochastic} composite RCS representing small-scale fading.
In the sequel,
we describe how these three factors are modeled and fitted.

\subsubsection{Mean Composite RCS}
As defined in~\eqref{eq:rcs_mult_new}, $\sigma_{\mathrm{M}}$ is the linear
mean of the composite RCS across all acquisitions of a given target,
thereby averaging over both geometry and small-scale fading
variability. It represents the average strength of the cluster. In the 3GPP notation of Table~\ref{tab:rcs_params_new},
$\sigma_{\mathrm{M}}$ is reported in dBsm for each target class.

\subsubsection{Deterministic Composite RCS}
The deterministic factor $\sigma_{\mathrm{D}}(\beta,\phi)$ captures the
systematic variation of the composite RCS with the bistatic angle
$\beta$ and the aspect angle $\phi$. The bistatic angle $\beta$
describes the relative geometry between the TX, the cluster center
(CC), and the RX, and is defined as the angle between the TX--CC
and RX--CC vectors,
\begin{equation}
\beta
=
\cos^{-1}\!\left(
\frac{
\mathbf{u}^{\mathrm{CC}}_{\mathrm{TX}}
\cdot
\mathbf{u}^{\mathrm{CC}}_{\mathrm{RX}}
}{
\|\mathbf{u}^{\mathrm{CC}}_{\mathrm{TX}}\|\,
\|\mathbf{u}^{\mathrm{CC}}_{\mathrm{RX}}\|
}
\right),
\label{eq:bistatic_angle}
\end{equation}
In contrast, the aspect angle $\phi$ parameterizes the directional
RCS response of the target. It is defined in the RX-centric
global coordinate system as
\begin{equation}
\phi
=
\cos^{-1}\!\left(
\frac{
\mathbf{u}^{\mathrm{CC}}_{\mathrm{RX}}
\cdot
\mathbf{u}^{\mathrm{RX}}
}{
\|\mathbf{u}^{\mathrm{CC}}_{\mathrm{RX}}\|
}
\right).
\label{eq:aspect_angle}
\end{equation}
Both angles are illustrated in Fig.~\ref{fig:GeometryModel}(b).

To isolate the target's directionality for modeling, each measured
composite RCS sweep $\sigma(\beta,\phi)$ is first circularly shifted by
$-\beta/2$, corresponding to the bistatic specular axis. This
operation aligns the dominant specular response with $\phi = 0^\circ$,
yielding a monostatic-equivalent angular response. The aligned sweeps
are then combined to obtain a single angular template,
$\tilde{\sigma}_{\mathrm{D}}(\phi)$, modeled using the 3GPP sector
function
\begin{equation}
\tilde{\sigma}_{\mathrm{D}}(\phi)\,(\mathrm{dB})
=
G_{\max}
-
\min\!\left\{
12\!\left(
\frac{\phi - \phi_{\mathrm{center}}}{\phi_{3\mathrm{dB}}}
\right)^{\!2},
\,\sigma_{\max}
\right\},
\label{eq:directional_pattern}
\end{equation}
followed by normalization to unit mean in the linear domain.
Because the composite RCS also decreases systematically with bistatic
angle $\beta$, each aligned sweep is normalized by the inverse of a
smooth bistatic weighting function
\begin{equation}
    w(\beta)
    =
    k_1
    \sin\!\left(
        \frac{k_2 \beta}{2}
    \right),
    \label{eq:bistatic_scaling}
\end{equation}
which parameterizes the deterministic bistatic dependence.

The deterministic factor is then reconstructed as the product of the
angular template and the bistatic weighting,
\begin{equation}
\sigma_{\mathrm{D}}(\beta,\phi)
=
w(\beta)\,
\tilde{\sigma}_{\mathrm{D}}\!\left(\phi + \frac{\beta}{2}\right),
\label{eq:sigmaD_reconstruct}
\end{equation}
where $\tilde{\sigma}_{\mathrm{D}}(\cdot)$ is understood in the linear
domain. All parameters in~\eqref{eq:sigmaD_reconstruct}, including
$G_{\max}$, $\phi_{\mathrm{center}}$, $\phi_{3\mathrm{dB}}$,
$\sigma_{\max}$, $k_1$, and $k_2$, are jointly optimized from the
measurement data.

Finally, since $\sigma_{\mathrm{D}}(\beta,\phi)$ appears multiplicatively
in~\eqref{eq:rcs_mult_new}, it is normalized to unit mean,
$\mathbb{E}\!\left[\sigma_{\mathrm{D}}(\beta,\phi)\right]=1$, to preserve
the mean composite RCS. Representative results for
$\tilde{\sigma}_{\mathrm{D}}(\phi)$ appear in
Fig.~\ref{fig:B1RadPattern_UAV}, and all fitted deterministic parameters
are listed in Table~\ref{tab:rcs_params_new}.

% ================== COMBINED RCS TABLE + FIGURES ==================
\begin{table*}[h!]
\centering
\caption{Composite RCS Model Parameters}
\label{tab:rcs_params_new}

\renewcommand{\arraystretch}{1.3}
\setlength{\tabcolsep}{6pt}

% ================== TABLE (TOP) ==================
\begin{minipage}{\textwidth}
\centering
\begin{tabular}{l|c|c|c|c}
\hline
\multirow{3}{*}{Target} &
\multirow{3}{*}{%
\begin{minipage}{0.10\textwidth}
\centering
$\sigma_{\mathrm{M}}$ (dBsm)\\
{\footnotesize (see (\ref{eq:rcs_mult_new}))}
\end{minipage}} &
\multicolumn{2}{c|}{Parameters of $\sigma_{\mathrm{D}}(\beta,\phi)$ (see (\ref{eq:sigmaD_reconstruct}))} &
\multirow{3}{*}{%
\begin{minipage}{0.07\textwidth}
\centering
$\varsigma_{\sigma_{\mathrm{S}}}$ (dB)\\
{\footnotesize (see (\ref{eq:lognormal_db_pdf}))}
\end{minipage}} \\
\cline{3-4}
 & &
Angular Pattern (dB) (see (\ref{eq:directional_pattern})) &
\shortstack{$k_1,\,k_2$\\{\footnotesize (see (\ref{eq:bistatic_scaling}))}} &
\\
\hline\hline

% UAV
UAV & -12.7 &
\begin{minipage}{0.50\textwidth}
\centering
\renewcommand{\arraystretch}{1.1}
\setlength{\tabcolsep}{3pt}
\begin{tabular}{l|c|c|c|c|c}
\hline
Sector & $\phi_{\text{center}}$ & $\phi_{\text{3dB}}$ & $G_{\max}$ & $\sigma_{\max}$ & $\phi$ Range \\ \hline\hline
Front  & 0°   & 13.36 & 6.28 & 9.63 & [315°,360°] U [0°,45°] \\ \hline
Left   & 95°  & 10.25 & 7.98 & 11.33 & [45°,135°] \\ \hline
Back   & 180° & 13.36 & 6.11 & 9.46 & [135°,225°] \\ \hline
Right  & 265° & 10.25 & 7.98 & 11.33 & [225°,315°] \\ \hline
\end{tabular}
\end{minipage}
& 3, 1.1 & 3.74 \\
\hline

% Robotic Arm
Robotic Arm & -5.14 &
\begin{minipage}{0.41\textwidth}
\centering
\renewcommand{\arraystretch}{1.1}
\setlength{\tabcolsep}{3pt}
\begin{tabular}{l|c|c|c|c|c}
\hline
Sector & $\phi_{\text{center}}$ & $\phi_{\text{3dB}}$ & $G_{\max}$ & $\sigma_{\max}$ & $\phi$ Range \\ \hline\hline
Arm & 90° & 50.44 & 0.95 & 9.48 & (40°,130°] U (230°,320°] \\ \hline
\end{tabular}
\end{minipage}
& 0.1, 0.1 & 4.61 \\
\hline

AMR             & 3.94  & 1 & -- & 6.77 \\ \hline
Human-1.60 m F  & -1.41 & 1 & -- & 8.20 \\ \hline
Human-1.69 m M  & 0.76  & 1 & -- & 8.12 \\ \hline
Human-1.85 m M  & 1.60  & 1 & -- & 7.84 \\ \hline

% Automobile-Sedan
Automobile-Sedan & 9.98 &
\begin{minipage}{0.50\textwidth}
\centering
\renewcommand{\arraystretch}{1.1}
\setlength{\tabcolsep}{3pt}
\begin{tabular}{l|c|c|c|c|c}
\hline
Sector & $\phi_{\text{center}}$ & $\phi_{\text{3dB}}$ & $G_{\max}$ & $\sigma_{\max}$ & $\phi$ Range  \\ \hline\hline
Front & 0° & 27.51 & -6.15 & 9.58 & [315°,360°] U [0°,45°] \\ \hline
Left  & 90° & 29.43 & 1.64  & 17.37 & [45°,135°] \\ \hline
Back  & 180° & 25.94 & -3.29 & 12.44 & [135°,225°] \\ \hline
Right & 270° & 29.43 & 1.64  & 17.37 & [225°,315°] \\ \hline
\end{tabular}
\end{minipage}
& 14, 0.5 & 6.40 \\
\hline

% SUV
Automobile-SUV & 13.36 &
\begin{minipage}{0.50\textwidth}
\centering
\renewcommand{\arraystretch}{1.1}
\setlength{\tabcolsep}{3pt}
\begin{tabular}{l|c|c|c|c|c}
\hline
Sector & $\phi_{\text{center}}$ & $\phi_{\text{3dB}}$ & $G_{\max}$ & $\sigma_{\max}$ & $\phi$ Range  \\ \hline\hline
Front & 0° & 31.53 & -2.03 & 13.7 & [315°,360°] U [0°,45°] \\ \hline
Left  & 90° & 32.46 & 6.76 & 22.49 & [45°,135°] \\ \hline
Back  & 180° & 30.95 & 0.83 & 16.56 & [135°,225°] \\ \hline
Right & 270° & 32.46 & 6.76 & 22.49 & [225°,315°] \\ \hline
\end{tabular}
\end{minipage}
& 15, 0.6 & 6.41 \\
\hline

% Van
Automobile-Van & 20.88 &
\begin{minipage}{0.50\textwidth}
\centering
\renewcommand{\arraystretch}{1.1}
\setlength{\tabcolsep}{3pt}
\begin{tabular}{l|c|c|c|c|c}
\hline
Sector & $\phi_{\text{center}}$ & $\phi_{\text{3dB}}$ & $G_{\max}$ & $\sigma_{\max}$ & $\phi$ Range \\ \hline\hline
Front & 0° & 30.43 & -2.25 & 13.48 & [315°,360°] U [0°,45°] \\ \hline
Left  & 90° & 32.82 & 6.54 & 22.27 & [45°,135°] \\ \hline
Back  & 180° & 30.92 & 0.61 & 16.34 & [135°,225°] \\ \hline
Right & 270° & 32.82 & 6.54 & 22.27 & [225°,315°] \\ \hline
\end{tabular}
\end{minipage}
& 15, 0.6 & 7.33 \\
\hline

\end{tabular}
\end{minipage}

\vspace{1.0em}

% ================== FIGURE 1 (MIDDLE) ==================
\begin{minipage}{\textwidth}
\centering
\setcounter{subtable}{0}

\subfloat[UAV]{%
    \includegraphics[width=0.20\textwidth]{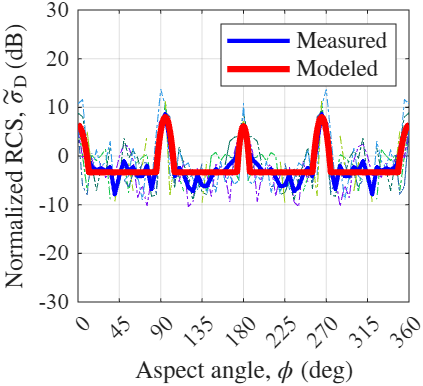}%
    \label{fig:B1RadPattern_UAV_sub}%
}
\hfill
\subfloat[Robotic Arm]{%
    \includegraphics[width=0.20\textwidth]{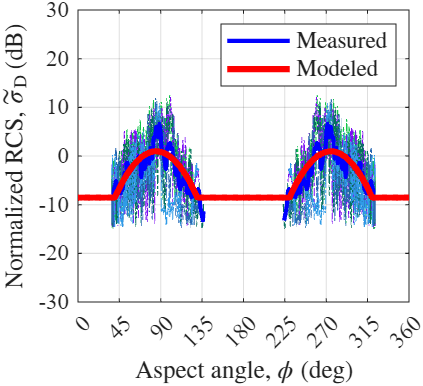}%
    \label{fig:B1RadPattern_Arm_sub}%
}
\hfill
\subfloat[AMR]{%
    \includegraphics[width=0.20\textwidth]{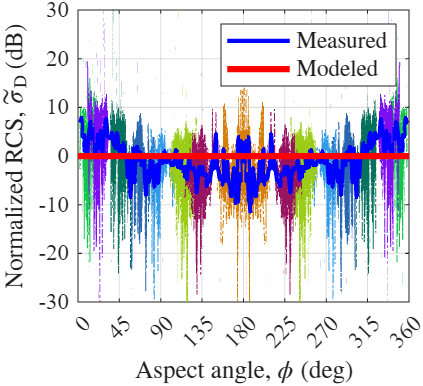}%
    \label{fig:B1RadPattern_AMR_sub}%
}
\hfill
\subfloat[Human--1.85\,m Male]{%
    \includegraphics[width=0.20\textwidth]{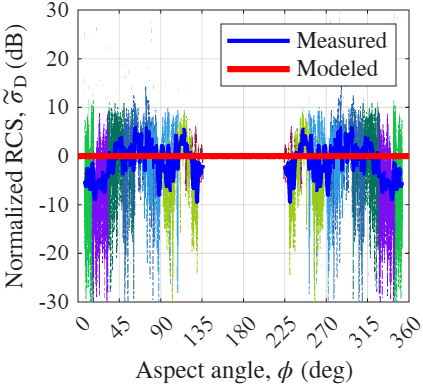}%
    \label{fig:B1RadPattern_Jihoon_sub}%
}
\hfill
\subfloat[Automobile--Van]{%
    \includegraphics[width=0.20\textwidth]{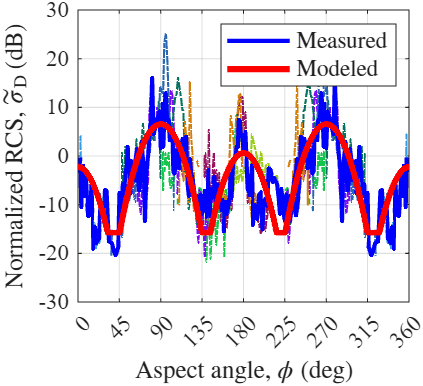}%
    \label{fig:B1RadPattern_Van_sub}%
}

\refstepcounter{figure}
\vspace{0.5em}

\raggedright
{Fig.~\thefigure. RCS model: Directional pattern fitting results for five representative targets. Thin colored curves show the measured normalized composite RCS $\tilde{\sigma}$ versus aspect angle $\phi$ after bistatic shifting by $-\beta/2$ and bistatic scaling by $k_1$ and $k_2$, with colors denoting different bistatic angles $\beta$. The thick blue curve shows the resulting average over $\beta$. The thick red curve shows the fitted directional pattern for the averaged response.}
\label{fig:B1RadPattern_UAV}
\end{minipage}

\vspace{1.0em}

% ================== FIGURE 2 (BOTTOM) ==================
\begin{minipage}{\textwidth}
\centering
\setcounter{subtable}{0}

\subfloat[UAV]{%
    \includegraphics[width=0.20\textwidth]{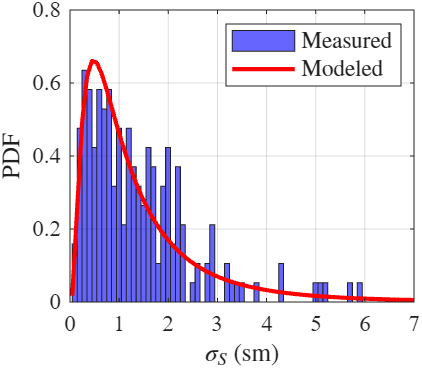}%
    \label{fig:B2Dist_UAV_sub}%
}
\hfill
\subfloat[Robotic Arm]{%
    \includegraphics[width=0.20\textwidth]{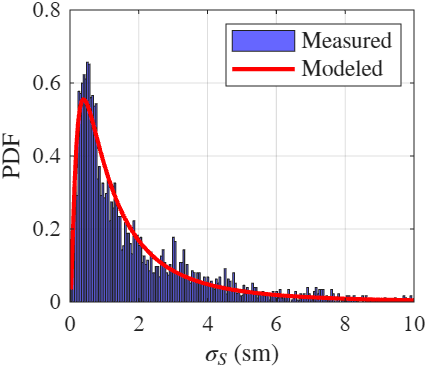}%
    \label{fig:B2Dist_Arm_sub}%
}
\hfill
\subfloat[AMR]{%
    \includegraphics[width=0.20\textwidth]{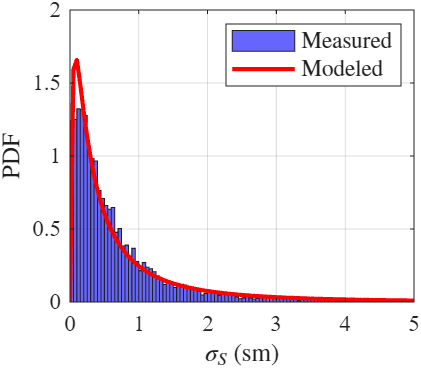}%
    \label{fig:B2Dist_AMR_sub}%
}
\hfill
\subfloat[Human--1.85\,m Male]{%
    \includegraphics[width=0.20\textwidth]{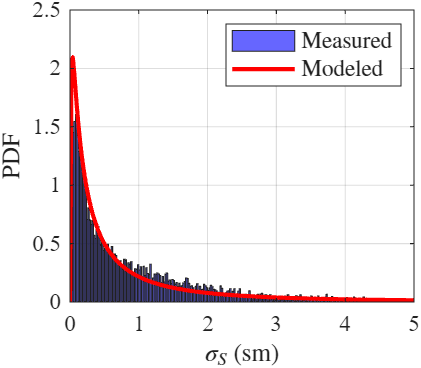}%
    \label{fig:B2Dist_Jihoon_sub}%
}
\hfill
\subfloat[Automobile--Van]{%
    \includegraphics[width=0.20\textwidth]{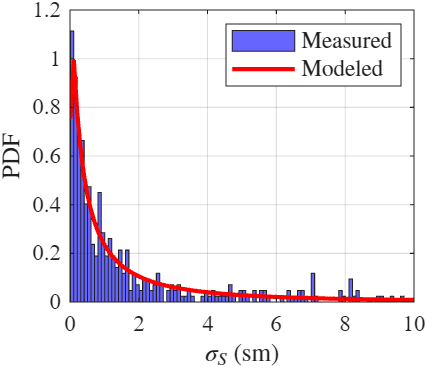}%
    \label{fig:B2Dist_Van_sub}%
}

\refstepcounter{figure}
\vspace{0.5em}

\raggedright
{\footnotesize Fig.~\thefigure. RCS model: Empirical and fitted distributions of the stochastic composite RCS $\sigma_{\mathrm{S}}$ for five representative targets.}
\label{fig:B2Dist}
\end{minipage}

\end{table*}
% ================== END COMBINED BLOCK ==================

\subsubsection{Stochastic Composite RCS}
The stochastic factor captures residual fluctuations in the composite
RCS after removing the mean factor $\sigma_{\mathrm{M}}$ and the
deterministic factor $\sigma_{\mathrm{D}}(\beta,\phi)$. For each
acquisition, the residual is defined as
\begin{equation}
\sigma_{\mathrm{S}}
=
\frac{\sigma(\beta,\phi)}
{\sigma_{\mathrm{M}}\,\sigma_{\mathrm{D}}(\beta,\phi)} .
\label{eq:small_scale_def}
\end{equation}
The stochastic term $\sigma_{\mathrm{S}}$ is modeled as a Lognormal random
variable with unit mean,
$\mathbb{E}\!\left[\sigma_{\mathrm{S}}\right]=1$, to ensure that it does
not bias the expected composite RCS under~\eqref{eq:rcs_mult_new}. In the
log-dB domain, the corresponding distribution is
\begin{equation}
f\!\left(10\log_{10}\sigma_{\mathrm{S}}\right)
=
\frac{1}{\varsigma_{\sigma_{\mathrm{S}}}\sqrt{2\pi}}
\exp\!\left(
-\frac{\left(10\log_{10}\sigma_{\mathrm{S}}\right)^2}
{2\,\varsigma_{\sigma_{\mathrm{S}}}^{2}}
\right),
\label{eq:lognormal_db_pdf}
\end{equation}
where $\varsigma_{\sigma_{\mathrm{S}}}$ denotes the standard deviation in
dB. Representative empirical distributions and fitted models are shown
in Fig.~\ref{fig:B2Dist}, and the extracted fluctuation spreads
$\varsigma_{\sigma_{\mathrm{S}}}$ for each target class are listed in
Table~\ref{tab:rcs_params_new}.

% ---------------------------------------------------------
\subsection{Incident-Point RCS}
\label{subsec:rcs_pathlevel_new}

The composite RCS in~\eqref{eq:rcs_mult_new} is formed by the coherent
superposition of the incident-point amplitudes
$\sqrt{\sigma_k}\,e^{j\varphi_k}$. Accordingly, the individual RCS values
$\sigma_k$ represent the power-contributing terms whose coherent sum
produces the total cluster strength. To statistically characterize the
distribution of point-level RCS values within a cluster, we introduce the
normalized quantity
\begin{equation}
\tilde{\sigma}
=
\frac{\sigma}{\sigma(\beta,\phi)},
\label{eq:relative_path_rcs}
\end{equation}
where $\sigma(\beta,\phi)$ denotes the composite RCS modeled in the
previous subsection. This normalization removes the absolute cluster
strength, isolating the dependence of individual point powers on the
underlying scattering geometry.

Measurements show that $\tilde{\sigma}$ decays systematically with the
distance offset $d^{\mathrm{CC}}$ from the cluster center. For a given
$d^{\mathrm{CC}}$, the empirical distribution of $\tilde{\sigma}$ is well
modeled by a Gaussian distribution in the linear domain,
\begin{equation}
f\!\left(\tilde{\sigma} \,\big|\, d^{\mathrm{CC}}\right)
=
\frac{1}{\varsigma_{\tilde{\sigma}}\!\left(d^{\mathrm{CC}}\right)\sqrt{2\pi}}
\exp\!\left(
-\frac{
\left(\tilde{\sigma}-\mu_{\tilde{\sigma}}\!\left(d^{\mathrm{CC}}\right)\right)^2
}{
2\,\varsigma_{\tilde{\sigma}}^{2}\!\left(d^{\mathrm{CC}}\right)
}
\right),
\label{eq:relative_path_gaussian}
\end{equation}
where both the mean and standard deviation depend on the distance offset
from the cluster center.
This dependence is parameterized per target class as
\begin{subequations}
\label{eq:mu_sigma_rk}
\begin{align}
\mu_{\tilde{\sigma}}\!\left(d^{\mathrm{CC}}\right)
&=
a_{\mu}\!\left(1 - e^{b_{\mu} d^{\mathrm{CC}}}\right),
\label{eq:mu_rk}
\\[4pt]
\varsigma_{\tilde{\sigma}}\!\left(d^{\mathrm{CC}}\right)
&=
a_{\varsigma}\!\left(1 - e^{b_{\varsigma} d^{\mathrm{CC}}}\right)
e^{c_{\varsigma} d^{\mathrm{CC}}},
\label{eq:sigma_rk}
\end{align}
\end{subequations}
where the parameters
$a_{\mu}$, $b_{\mu}$, $a_{\varsigma}$, $b_{\varsigma}$, and
$c_{\varsigma}$ are fitted independently for each target class. These
functions capture both the rapid decay of dominant points from the
cluster center and the increasing variability of weaker points at larger
distance offsets. Representative fits of
$\mu_{\tilde{\sigma}}(d^{\mathrm{CC}})$ and
$\varsigma_{\tilde{\sigma}}(d^{\mathrm{CC}})$ for the different target
classes are shown in Fig.~\ref{fig:RelativeRCSmodel}, and the
corresponding parameters are listed in
Table~\ref{table:RelativeRCSmodel}.

The phases $\varphi_k$ were found empirically to be well modeled as
independent and uniformly distributed over $[0,2\pi)$, which is expected
at 28~GHz: the wavelength is short relative to the separations between
resolvable points. This geometric
separation decorrelates their phase differences, producing the observed
uniform distribution.

% ---------- TABLE + FIGURE (combined) ----------
\begin{table*}[h!]
\centering
\caption{Incident-Point RCS Model Parameters}
\label{table:RelativeRCSmodel}
\renewcommand{\arraystretch}{1.2}
\setlength{\tabcolsep}{8pt}

% ---------- TOP: PARAMETER TABLE ----------
\begin{minipage}{\textwidth}
\centering
\begin{tabular}{l|c|c|c|c|c}
\hline
\multirow{2}{*}{Target} &
\multicolumn{2}{c|}{$\mu_{\tilde{\sigma}}(d_k^{\mathrm{CC}})$ (see (\ref{eq:mu_sigma_rk}a))} &
\multicolumn{3}{c}{$\varsigma_{\tilde{\sigma}}(d_k^{\mathrm{CC}})$ (see (\ref{eq:mu_sigma_rk}b))} \\
\cline{2-6}
& $a_{\mu}$ & $b_{\mu}$ & $a_{\varsigma}$ & $b_{\varsigma}$ & $c_{\varsigma}$ \\
\hline\hline
UAV                & -14.9 & -3.0 & 4.8  & -5.7  & 0     \\
Robotic Arm        & -15.0 & -4.4 & 42.2 & -1.3  & -4.4  \\
AMR                & -17.6 & -2.4 & 5.1  & -17.4 & -0.1  \\
Human--1.60\,m F   & -15.9 & -3.7 & 12.3 & -4.0  & -1.0  \\
Human--1.69\,m M   & -24.1 & -2.8 & 10.2 & -6.0  & -0.1  \\
Human--1.85\,m M   & -16.3 & -4.7 & 6.8  & -25.1 & 0     \\
Automobile--Sedan  & -9.6  & -3.7 & 21.6 & -1.3  & -1.0  \\
Automobile--SUV    & -12.7 & -4.7 & 6.8  & -25.1 & 0     \\
Automobile--Van    & -16.3 & -2.9 & 9.5  & -8.5  & -0.1  \\
\hline
\end{tabular}
\end{minipage}

\vspace{1em}

% ---------- BOTTOM: PLOTS ----------
\begin{minipage}{\textwidth}
\centering

\subfloat[UAV]{\includegraphics[width=0.19\textwidth]{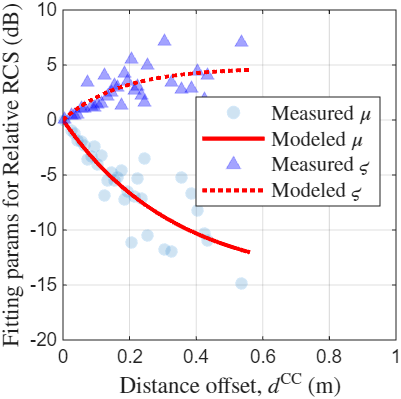}}
\hfill
\subfloat[Robotic Arm]{\includegraphics[width=0.19\textwidth]{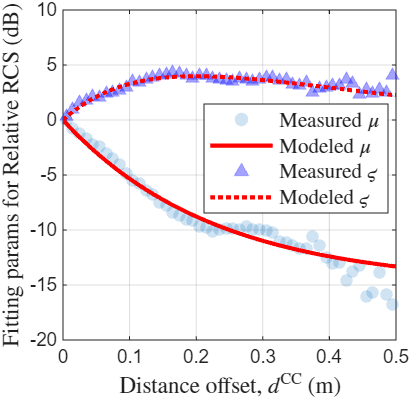}}
\hfill
\subfloat[AMR]{\includegraphics[width=0.19\textwidth]{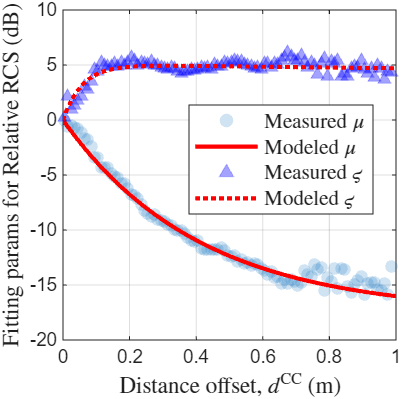}}
\hfill
\subfloat[Human--1.85\,m Male]{\includegraphics[width=0.19\textwidth]{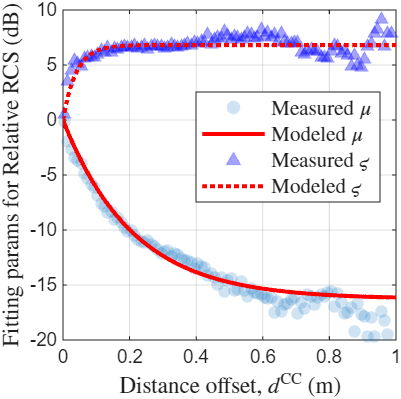}}
\hfill
\subfloat[Automobile--Van]{\includegraphics[width=0.19\textwidth]{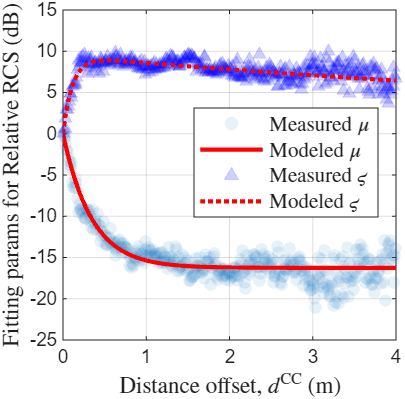}}

% ---------- MANUAL FIGURE CAPTION WITH NUMBER ----------
\refstepcounter{figure}
\vspace{0.5em}
\raggedright
{ Fig.~\thefigure. RCS model: incident-point RCS parameters' trends versus distance offset for five representative targets.}
\label{fig:RelativeRCSmodel}
\end{minipage}
\end{table*}

\subsection{Parameter Analysis}
\label{subsec:rcs_parameter_analysis}

The fitted RCS parameters in
Tables~\ref{tab:rcs_params_new} and~\ref{table:RelativeRCSmodel},
together with the empirical distributions in
Figs.~\ref{fig:B1RadPattern_UAV}--\ref{fig:RelativeRCSmodel},
exhibit clear and physically interpretable trends across the nine
measured targets. These trends reflect systematic differences in target
size, material composition, facet structure, and geometric complexity,
and collectively validate the proposed decomposition of composite and
incident-point-level RCS.

\vspace{0.3em}\noindent
\textit{Mean Composite RCS:}
The mean composite RCS $\sigma_{\mathrm{M}}$ increases monotonically with
overall target size and illuminated surface extent. Human subjects follow
a strict ordering with height, while the three automobiles exhibit the
expected progression sedan $<$ SUV $<$ van. The AMR and robotic arm
occupy intermediate values consistent with their dimensions and metallic
construction. 
For reference, we compare our mean composite RCS values,
$\sigma_{\mathrm{M}}$, with published RCS values near 28~GHz for similar
target classes, while noting differences in target construction as well as
measurement frequency, polarization, and geometry. For UAVs, reported values are $-9.6$~dBsm for a mid-size UAV
\cite{azim20253gpp} and $-9.5$~dBsm for a hexacopter
\cite{semkin2020analyzing}, compared with $-12.7$~dBsm for our UAV. For robotic platforms, reported values are $-8.165$~dBsm
for a robotic arm and $-11.235$~dBsm for an automated guided vehicle
(AGV)-type mobile robot \cite{azim20253gpp}, compared with $-5.14$~dBsm
for our robotic arm and $3.94$~dBsm for our taller instrumented AMR, respectively. For human targets, prior measurements report a value
of $-3.78$~dBsm \cite{manfredi2020efficient}, values ranging from
$-5.47$~dBsm to $-2.19$~dBsm across front, rear, and side views
\cite{fortuny2013radar}, and pedestrian extrema spanning $-15$~dBsm to
$5$~dBsm \cite{nataraja2025cars}. These values are consistent in scale
with the values of $-1.41$~dBsm, $0.76$~dBsm, and $1.60$~dBsm obtained
for our 1.60~m, 1.69~m, and 1.85~m subjects, respectively. For vehicles,
reported sedan extrema range from $-15$~dBsm to $25$~dBsm
\cite{nataraja2025cars}, bracketing the values of $9.98$~dBsm,
$13.36$~dBsm, and $20.88$~dBsm obtained for our sedan, SUV, and van,
respectively.

\vspace{0.3em}\noindent
\textit{Deterministic Composite RCS:}
The deterministic angular dependence $\sigma_{\mathrm{D}}(\beta,\phi)$,
illustrated in Fig.~\ref{fig:B1RadPattern_UAV} and parameterized in
Table~\ref{tab:rcs_params_new}, also follows intuitive structure.
Targets with stable, oriented facets—UAV, robotic arm, and all three
automobiles—exhibit pronounced directional lobes, whereas humans and the
AMR, which lack persistent planar features, show comparatively weak
angular variation. The fitted 3GPP sector patterns recover these
differences naturally: automobile responses are broad and high-gain, while
human and AMR patterns reduce toward the isotropic case. The dependence
on bistatic geometry is governed by the scaling parameters $k_1$ and
$k_2$, which modulate the angular response as the TX--CC--RX configuration
varies.

\vspace{0.3em}\noindent
\textit{Stochastic Composite RCS:}
Examination of the stochastic fluctuation parameter
$\varsigma_{\sigma_{\mathrm{S}}}$ in Fig.~\ref{fig:B2Dist} and
Table~\ref{tab:rcs_params_new} reveals a consistent cross-target trend.
Targets with strong deterministic directionality generally exhibit
smaller residual fluctuation variance after removal of the mean and
deterministic factors. In these cases, $\sigma_{\mathrm{D}}(\beta,\phi)$
absorbs much of the geometry-driven variation, leaving less randomness
to $\sigma_{\mathrm{S}}$. Human subjects, which are closer to isotropic
and exhibit rapidly varying local geometry, show the largest
$\varsigma_{\sigma_{\mathrm{S}}}$ values. The UAV, AMR, and automobiles
exhibit lower fluctuation levels, reflecting more stable scattering
structure. The van constitutes a notable exception: despite strong
directionality, its large physical extent introduces many independent
scattering regions whose relative contributions vary across
acquisitions, resulting in elevated residual variance. Literature values
reported near 28~GHz for $\varsigma_{\sigma_{\mathrm{S}}}$ include $10.66$~dB for a mid-size UAV, $13.54$~dB for a robotic arm, and $6.27$~dB for an AGV \cite{azim20253gpp}, compared with $3.74$~dB, $4.61$~dB, and
$6.77$~dB for our UAV, robotic arm, and AMR, respectively.

\vspace{0.3em}\noindent
\textit{Incident-Point RCS:}
The RCS redistribution parameters in
Table~\ref{table:RelativeRCSmodel} and
Fig.~\ref{fig:RelativeRCSmodel} elucidate how the composite RCS is
distributed across the incident points. Larger targets
such as automobiles and the AMR exhibit slower decay of
$\mu_{\tilde{\sigma}}(d^{\mathrm{CC}})$, consistent with extended
illuminated surfaces that support strong near-specular interactions over
a wider region. More articulated or irregular targets, including humans
and the robotic arm, exhibit larger spreads
$\varsigma_{\tilde{\sigma}}(d^{\mathrm{CC}})$, reflecting increased
variability in how local facets contribute to individual point powers.
Compact platforms such as the UAV show both rapid decay and small
variance, consistent with a localized scattering region.
% Requires \usepackage{subfig}

\section{Channel Realization}
\label{sec:channel_realization}
This section describes the algorithm used to generate channel
realizations from the reported target-specific channel parameters. The
resulting realizations represent the target-channel component of a
complete ISAC scene. Environmental clutter is not modeled as part of the
target-specific scatter model; instead, it can be generated as a
separate background channel and superposed with one or more target
channels, consistent with the 3GPP TR~38.901 channel
model~\cite{3gpp38901}. The first subsection outlines the target-channel
realization procedure, while the second presents representative
realizations.

\subsection{Channel Realization Procedure}
The instantaneous channel realization at time $t$ is constructed by combining the geometric scattering framework of
Section~\ref{sec:cm_geom} with the RCS model of
Section~\ref{sec:cm_rcs}. As illustrated in
Fig.~\ref{fig:realization_hierarchy}, the realization proceeds across
three hierarchical levels: 1) the cluster level; 2) the incident-point level; and 3) the path level. The green blocks denote the geometric model parameter inputs, the yellow blocks the stochastic model parameter inputs, and the red blocks the intermediate level inputs/outputs.

\vspace{1em}
\subsubsection*{1) Cluster Level}

\begin{figure}[!t]
\includegraphics[width=\columnwidth]{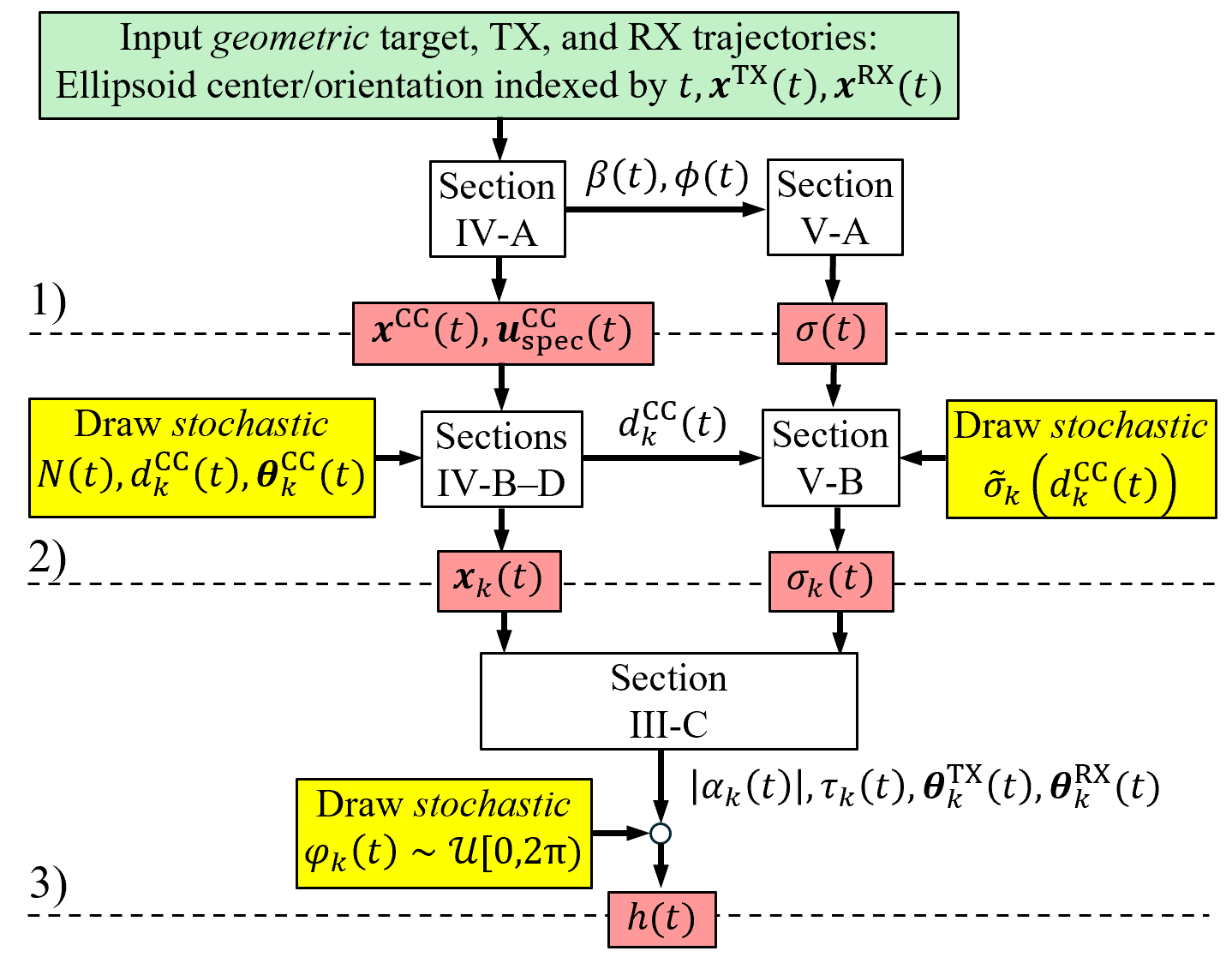}
\caption{Hierarchical channel realization flowchart. 1) Cluster-level realization. 2) Incident-point-level realization. 3) Path-level realization.}
\label{fig:realization_hierarchy}
\end{figure}

As illustrated in Fig.~\ref{fig:realization_hierarchy}, the cluster
level determines the cluster center and bistatic specular axis and the composite RCS. 

\vspace{0.5em}
\noindent
\textit{Cluster geometry:}
In the general case, the target is modeled as a canonical ellipsoid at
time $t$, consistent with Section~\ref{subsec:sc_definition}, which
uniquely determines the bistatic--specular cluster center
$\mathbf{x}^{\mathrm{CC}}(t)$ and the associated bistatic specular axis
$\mathbf{u}^{CC}_{\mathrm{spec}}(t)$ from Section~\ref{subsec:sc_definition}. 
For nondirectional targets, the ellipsoid may be replaced by
a sphere and, in the limiting case of a compact target (negligible
sphere radius), by a point scatterer.

\vspace{0.5em}
\noindent
\textit{Composite RCS:}
Given the instantaneous geometry, the composite RCS $\sigma(t)$ is
evaluated using the multiplicative model in
\eqref{eq:rcs_mult_new}. The required bistatic angle $\beta(t)$ and
aspect angle $\phi(t)$ follow directly from the instantaneous geometry
via \eqref{eq:bistatic_angle} and \eqref{eq:aspect_angle}, and the
deterministic and stochastic factors are applied as described in
Section~\ref{subsec:composite_rcs}.

\vspace{1em}
\subsubsection*{2) Incident-Point Level}

As illustrated in Fig.~\ref{fig:realization_hierarchy}, the
incident-point level stochastically instantiates $N(t)$ incident points
around the cluster geometry and redistributes the composite RCS across
them. The number of paths $N(t)$ associated with the target is drawn
according to the Poisson distribution in
\eqref{eq:distribution path count} in Section~\ref{subsec:path count}.

\vspace{0.5em}
\noindent
\textit{Incident-point geometry:}
For each path $k=1,\dots,N(t)$, an incident point $\mathbf{x}_k(t)$ is
generated relative to $\mathbf{x}^{\mathrm{CC}}(t)$ by
\begin{itemize}
\item drawing the distance offset $d_k^{\mathrm{CC}}(t)$ according to the
Gamma distribution in~\eqref{eq:dist_dsc} from Section~\ref{subsec:distance offset};
\item drawing the elevation angle offset
$\theta_k^{\mathrm{CC,E}}(t)$ according to the exponential distribution
in~\eqref{eq:dist_ang_offset} from Section~\ref{subsec:angle offset};
\item drawing the azimuthal angle offset
$\theta_k^{\mathrm{CC,A}}(t)$ uniformly over $[0^\circ,360^\circ)$.
\end{itemize}
These parameters uniquely determine the 3D incident point
$\mathbf{x}_k(t)$ of each path.

\vspace{0.5em}
\noindent
\textit{Incident-point RCS:}
The composite RCS is redistributed across the $N(t)$ incident points
using the path-level RCS model in Section~\ref{subsec:rcs_pathlevel_new}.
For each path, a normalized RCS $\tilde{\sigma}_k(t)$ is drawn
conditionally on $d_k^{\mathrm{CC}}(t)$ according to
\eqref{eq:relative_path_gaussian}--\eqref{eq:mu_sigma_rk}, and the
absolute path RCS is obtained as
\(
\sigma_k(t)=\tilde{\sigma}_k(t)\,\sigma(t).
\)

\vspace{1em}
\subsubsection*{3) Path Level}

As illustrated in Fig.~\ref{fig:realization_hierarchy}, each realized
incident point and its associated RCS map to a set of path properties,
which correspond to the inverse of the path-to--incident-point mapping
in Section~\ref{subsec:path_rcs}. These properties are then used to
synthesize the instantaneous channel response $h(t)$.

\vspace{0.5em}
\noindent
\textit{Path geometry:}
For each incident point $\mathbf{x}_k(t)$, the geometric path
parameters follow directly from the TX and RX distances to
$\mathbf{x}_k(t)$ given by $d_k^{\mathrm{TX}}(t)$ and
$d_k^{\mathrm{RX}}(t)$ in \eqref{eq:tx_rx_distance_def}. Their sum
defines the total bistatic path length
$d_k(t)=d_k^{\mathrm{TX}}(t)+d_k^{\mathrm{RX}}(t)$, which satisfies
the bistatic range condition \eqref{eq:bistatic_range_constraint}
and yields the propagation delay $\tau_k(t)=d_k(t)/c$. Under the
first-order interaction assumption, the incident point also uniquely
determines the AoD at the TX and the AoA at the RX. Let
$\mathbf{x}=[x,y,z]^{\mathsf T}$ denote a generic Cartesian position
vector. The AoD is computed as
\[
\begin{aligned}
\theta_k^{\mathrm{TX,A}}(t)
&=
\tan^{-1}\!\left(
\dfrac{y_k(t)-y^{\mathrm{TX}}(t)}
{x_k(t)-x^{\mathrm{TX}}(t)}
\right)\\[4pt]
\theta_k^{\mathrm{TX,E}}(t)
&=
\sin^{-1}\!\left(
\dfrac{z_k(t)-z^{\mathrm{TX}}(t)}
{\|\mathbf{x}_k(t)-\mathbf{x}^{\mathrm{TX}}(t)\|}
\right)
\end{aligned}
\]
and the AoA as
\[
\begin{aligned}
\theta_k^{\mathrm{RX,A}}(t)
&=
\tan^{-1}\!\left(
\dfrac{y_k(t)-y^{\mathrm{RX}}(t)}
{x_k(t)-x^{\mathrm{RX}}(t)}
\right)\\[4pt]
\theta_k^{\mathrm{RX,E}}(t)
&=
\sin^{-1}\!\left(
\dfrac{z_k(t)-z^{\mathrm{RX}}(t)}
{\|\mathbf{x}_k(t)-\mathbf{x}^{\mathrm{RX}}(t)\|}
\right)
\end{aligned}
\]

\vspace{0.5em}
\noindent
\textit{Path complex amplitude:}
Each path RCS $\sigma_k(t)$ is then mapped to the magnitude of the
complex path coefficient $|\alpha_k(t)|$ through the bistatic radar equation  in~\eqref{eq:path_radar_equation}.
An independent random phase
$\varphi_k(t)\sim\mathcal{U}[0,2\pi)$ is assigned, yielding
\(
\alpha_k(t)=|\alpha_k(t)|e^{j\varphi_k(t)}.
\)

\vspace{0.5em}
\noindent
\textit{Channel synthesis:}
Given the complete set of path properties at time $t$, the instantaneous
channel response is synthesized in the joint angle--delay domain as
\begin{IEEEeqnarray}{rCl}
h(t)
&=&
\sum_{k=1}^{N(t)}
\alpha_k(t)\,
e^{j2\pi f_c \tau_k(t)} \label{eq:ht}
\\
&\cdot&
\delta\!\left(\tau-\tau_k(t),
\boldsymbol{\theta}^{\mathrm{TX}}
-
\boldsymbol{\theta}^{\mathrm{TX}}_k(t),
\boldsymbol{\theta}^{\mathrm{RX}}
-
\boldsymbol{\theta}^{\mathrm{RX}}_k(t)
\right). \nonumber
\end{IEEEeqnarray}

\vspace{0.5em}
\noindent
\textit{Pathloss:}
The corresponding instantaneous pathloss is defined as
\begin{equation}
PL(t)=\frac{1}{|h(t)|^2}.
\label{eq:PLt}
\end{equation}

\begin{figure*}[ht!] 
\centering 
\subfloat[UAV]{ 
\includegraphics[width=0.19\textwidth]{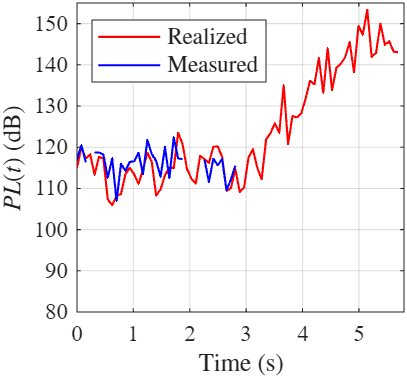}} \hfill \subfloat[Robotic Arm]{ 
\includegraphics[width=0.19\textwidth]{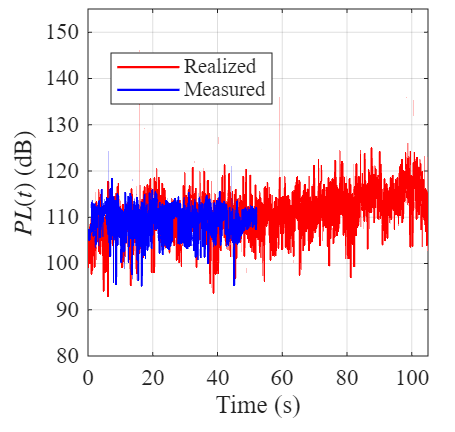}} \hfill \subfloat[AMR]{\includegraphics[width=0.195\textwidth]{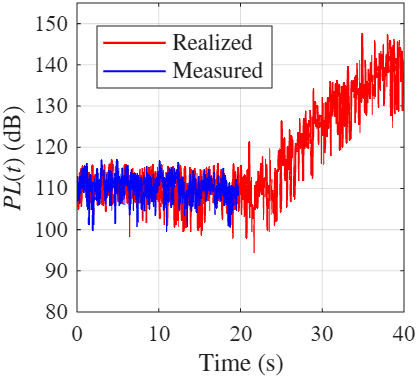}} \hfill 
\subfloat[Human--1.85\,m Male]{ \includegraphics[width=0.19\textwidth]{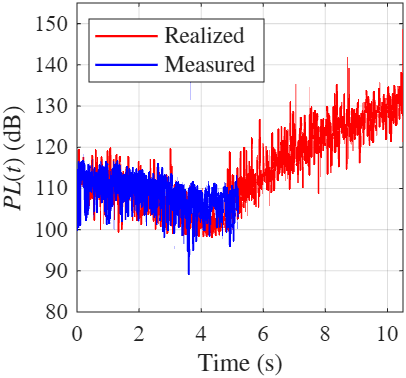}} \hfill 
\subfloat[Automobile--Van]{\includegraphics[width=0.2\textwidth]{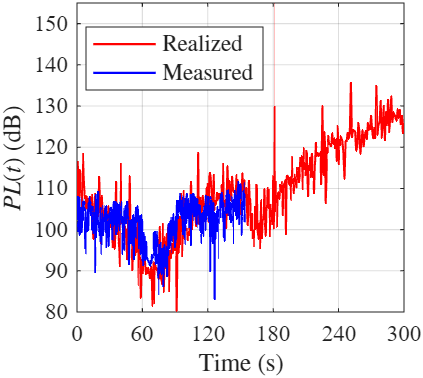}} 

% ---------- MANUAL FIGURE CAPTION WITH NUMBER ---------- 
\refstepcounter{figure} \vspace{0.5em} \raggedright {\footnotesize Fig. ~\thefigure. Realized versus measured pathloss for five representative targets.} \label{fig:PLcomparison} 
\end{figure*}

\subsection{Example Realizations}
A software implementation written in MATLAB\footnotemark[2] to realize the channel model for all targets is available here~\cite{nist_target_channel_software}.
Fig.~\ref{fig:PLcomparison} compares the measured and realized pathloss
evolution for five representative target classes under dynamic motion.
In all cases, both measured and realized pathloss are computed
identically using the channel synthesis and pathloss definitions
in~\eqref{eq:ht} and~\eqref{eq:PLt}, i.e., by coherently summing all
resolved target-associated paths and taking the inverse of the resulting
channel power gain. For the first half of the observation window, the
realized trajectories exactly replicate the measured motion, where
acquisitions are indexed by the time variable $t$; this interval
therefore serves as a direct validation of the proposed geometric and
RCS-based model under observed dynamics. During the second half,
controlled trajectory variations are introduced that deviate from the
measured motion, for extrapolation beyond the measurement conditions.

For the UAV case, the first half corresponds to a continuous rotation
from 0$^\circ$ to 180$^\circ$ at a fixed location, capturing
aspect-dependent RCS variations; the resulting pathloss minima align with
measured RCS peaks at favorable aspect angles. In the second half, the
UAV is assumed to recede from the TX and RX at constant speed, producing
the increasing pathloss trend toward the end of the interval. For the robotic arm, the realized motion follows the measured articulated
trajectory during the first half, while in the second half the arm
configuration is driven away from the measured trajectory to the extent
permitted by the arm reach, producing a gradual pathloss increase under
physically feasible articulated motion.
For the human and AMR cases, the realized trajectories follow the
measured approach motion during the first half, resulting in reduced
pathloss due to decreasing bistatic separation and improved geometric
alignment. In the second half, both targets move away from the TX and
RX, leading to a gradual pathloss increase consistent with increasing
propagation distance. For the van case, the TX follows the
measured motion initially, after which it remains stationary while the
automobile recedes from both TX and RX, producing a monotonic pathloss
increase; the pathloss reduction observed while the TX is moving
coincides with measured RCS peaks from the automobile side.

Overall, the close agreement in pathloss trends and fluctuation
statistics across both validation and extrapolation intervals
demonstrates that the proposed target-aware channel modeling framework
captures motion-induced pathloss dynamics across diverse target classes.

\section{Conclusion}
\label{sec:conclusion}
This paper presented a unified channel model for scattering from representative ISAC targets based on an exhaustive campaign of 28~GHz channel measurements.
By combining a 256-element phased-array receiver with super-resolution
processing, multiple paths were reliably extracted across
all target classes, enabling a path-specific view of target-induced
scattering that is directly tied to physical geometry.
Target scattering was shown to be well represented by clusters of paths
anchored at a bistatic-specular cluster center defined by the
TX--target--RX geometry, providing an explicit and interpretable link
between propagation geometry and target shape. A deterministic geometric
backbone ensures spatial--temporal consistency as targets move and reorient,
while a stochastic description tuned directly from measurements captures
the observed variability in scattering behavior without sacrificing
physical intuition.

Across nine heterogeneous targets, the resulting model parameters exhibit
consistent and interpretable trends: larger and more extended targets
produce broader clusters, while material composition and geometric
complexity introduce systematic variations in angular spread and power
fluctuations. The inferred statistical behavior is well behaved and stable
across geometries and motion, supporting robust channel realization.
Validation through synthesized channel realizations demonstrates close
agreement with measured pathloss evolution under dynamic motion, confirming
that the proposed model faithfully reproduces both large-scale trends and
path-level fluctuations. Together, these results establish a
path-specific, geometry-driven, and measurement-tuned channel
representation that preserves target shape information while remaining
compatible with communication-oriented abstractions. As such, the proposed
model is well suited for ISAC network simulation and evaluation.

\appendices
\section{}
\label{app:near_field}
\begin{figure}[!h]
    \centering
    \includegraphics[width=\linewidth]
    {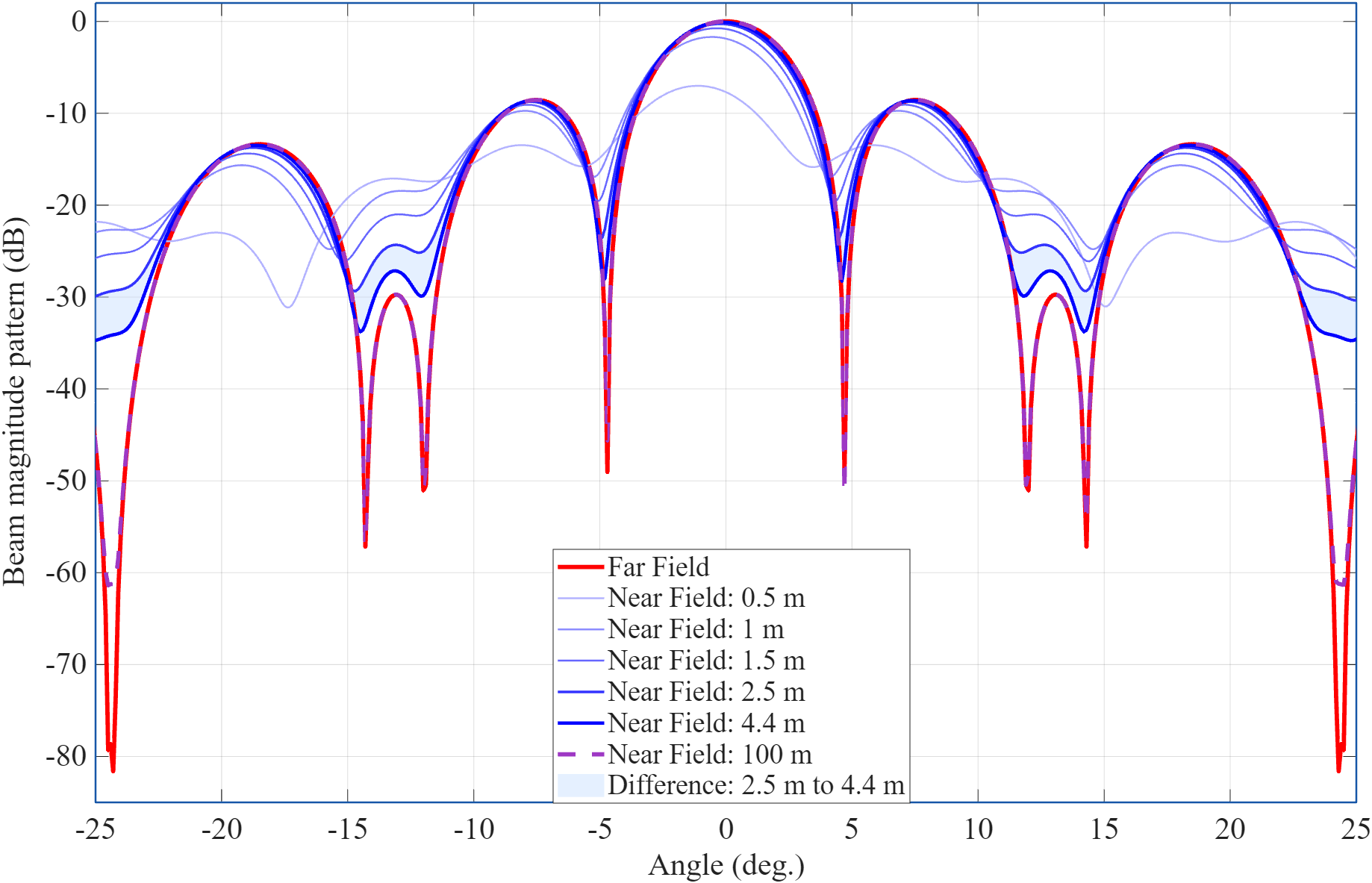}
    \par\vspace{1em}
    \includegraphics[width=\linewidth]
    {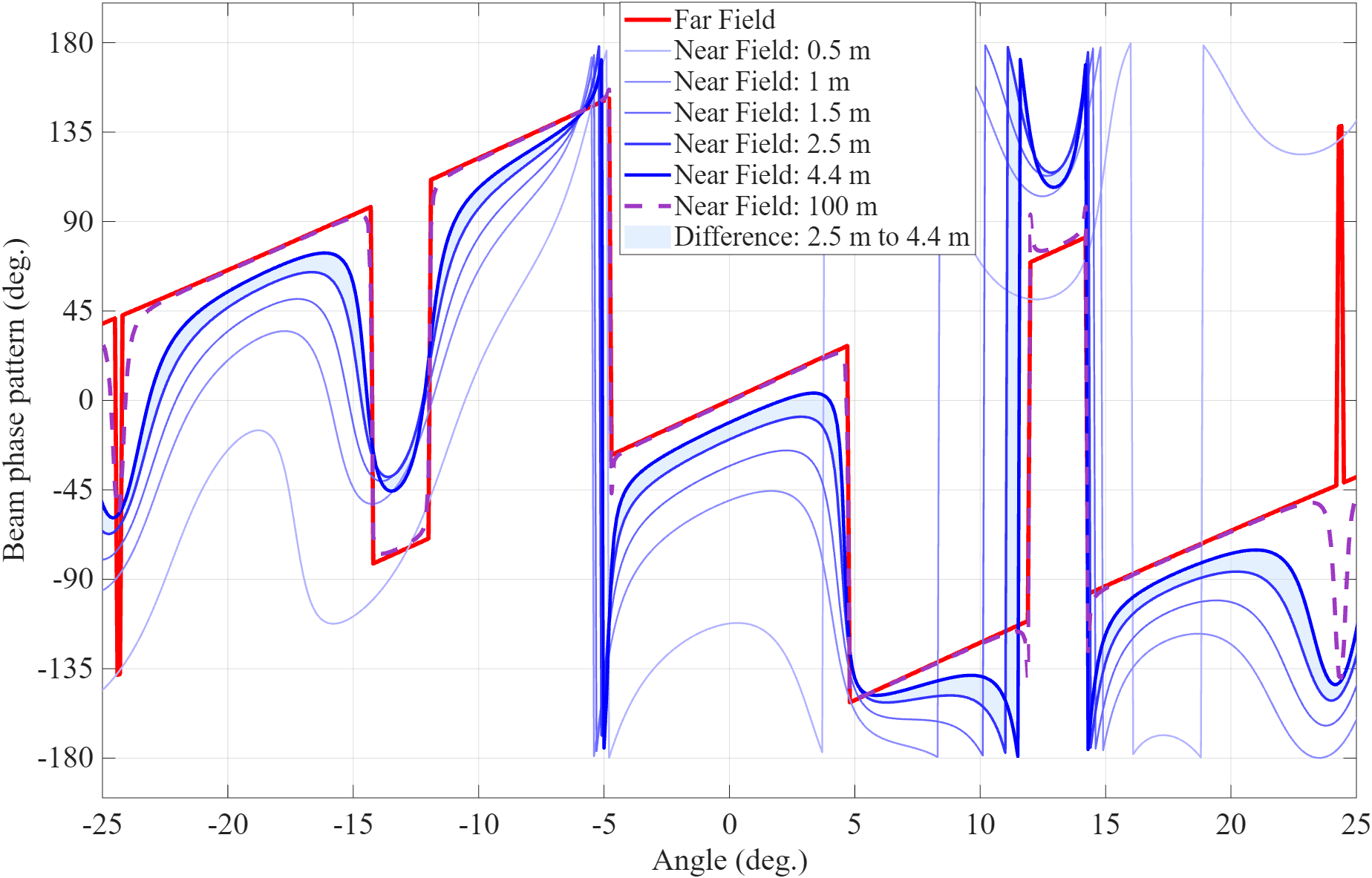}
    \caption{Far-field and near-field RX beamforming patterns at
    several scatterer distances: (a) normalized magnitude beam pattern and
    (b) referenced beam phase pattern. The shaded region highlights the
    difference between the patterns at the minimum measurement distance of
    2.5~m and the 4.4~m Fraunhofer distance.}
    \label{fig:nf_ff_validation}
\end{figure}

Here we validate the far-field planar-wave assumption for our RX array. At 28.5~GHz, the Fraunhofer distance
$2D^2/\lambda$---the conventional far-field threshold corresponding to a
maximum center-to-edge phase deviation of $22.5^\circ$---is 4.4~m for the
RX aperture ($D=152.4$~mm). However, the minimum RX--target distance in the
measurement campaign is 2.5~m and thus lies within the near-field region,
for which the corresponding phase deviation is approximately $40^\circ$.
To quantify the effect of spherical-wave curvature within the near-field
using the actual RX element geometry, spherical wavefronts were generated
and processed using the array's far-field planar-wave beamforming weights. Fig.~\ref{fig:nf_ff_validation}
shows the magnitude and phase cuts of the beam pattern at azimuth
$0^\circ$ for several finite scatterer distances, together with the
far-field pattern. The complementary cut at elevation $0^\circ$ was also
evaluated and exhibited similar behavior.

The magnitude plots in
Fig.~\ref{fig:nf_ff_validation}(a) show rapid convergence of the
near-field pattern toward the far-field. The pattern at 2.5~m
remains close to the 4.4~m Fraunhofer pattern (as well as the far-field
pattern), particularly in the main-lobe region, whereas the most pronounced
deviations occur below $-23$~dB, far
from the main peak. The shaded region in
Fig.~\ref{fig:nf_ff_validation}(a) highlights the difference between the
2.5~m and 4.4~m patterns. The normalized root-mean-square error (NRMSE)
with respect to the far-field pattern, computed in linear magnitude
over the full angular cut, is 9.46\,\% at 2.5~m. The error decreases
further to 5.15\,\% at 4.4~m, mainly due to reduced discrepancies in the
low-level sidelobes. Because the magnitude patterns in
Fig.~\ref{fig:nf_ff_validation}(a) are normalized by the same far-field
peak, any array-gain defocusing, quantified as near-field peak-gain
reduction, remains visible. For a scatterer at the azimuth
$0^\circ$ steering direction, the gain at 2.5~m is
$0.19$~dB below the far-field value, which is within the
0.48~dB path-gain error reported for the measurement system
in~\cite{caudill2021real}. Therefore, the array-gain defocusing causes a negligible estimation error in the scatterer's path gain or RCS.

The phase plots in Fig.~\ref{fig:nf_ff_validation}(b) also show rapid convergence  of the near-field pattern toward the far-field reference. However, even at 100~m, more than an order of
magnitude beyond the 4.4~m Fraunhofer distance, the near-field pattern
does not reach the ideal far-field. This suggests that the Fraunhofer
distance should be interpreted as a guideline rather
than a sharp boundary. The residual phase difference between 2.5~m and
4.4~m, shaded in Fig.~\ref{fig:nf_ff_validation}(b), remains below
$7^\circ$ over the angular region considered here, comparable to the approximately $6^\circ$ standard deviation of the measurement system's phase noise
reported in~\cite{caudill2021real}. Thus, the phase deviation introduced
by using planar-wave beamforming at 2.5~m is on the order of the system
phase uncertainty.

\section{}
\label{app:path_validation}
\begin{figure}[!h]
    \centering
    \includegraphics[width=0.49\linewidth]
    {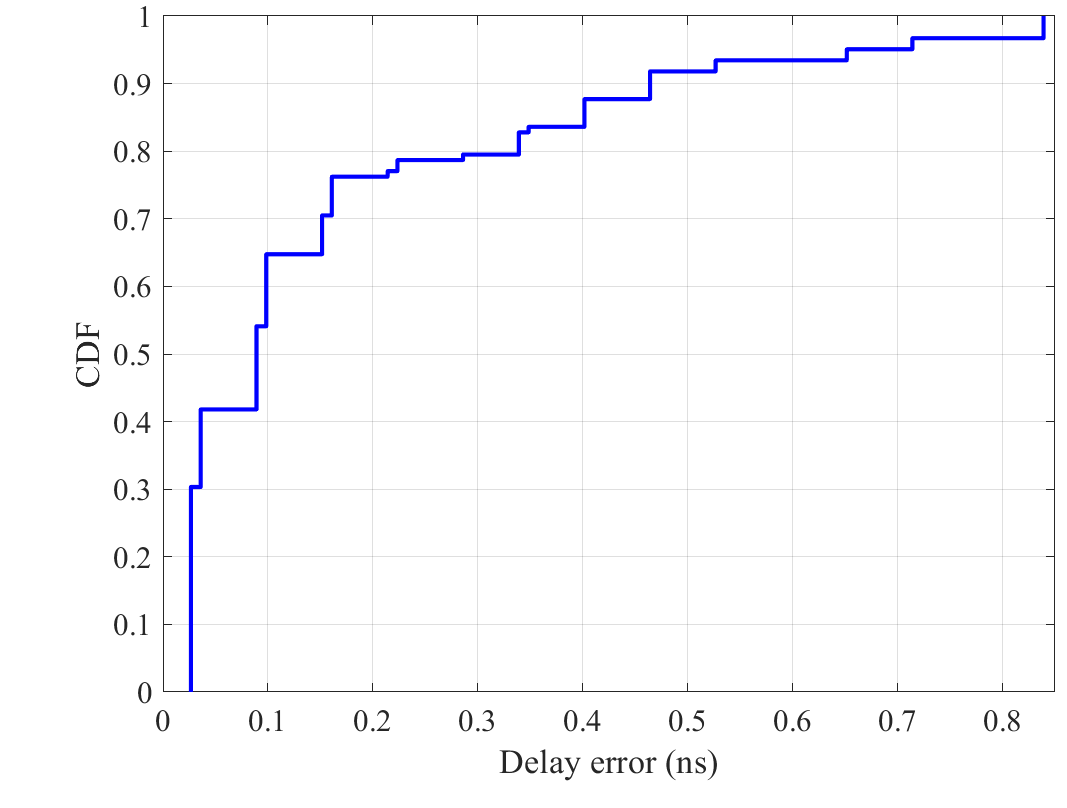}
    \includegraphics[width=0.49\linewidth]
    {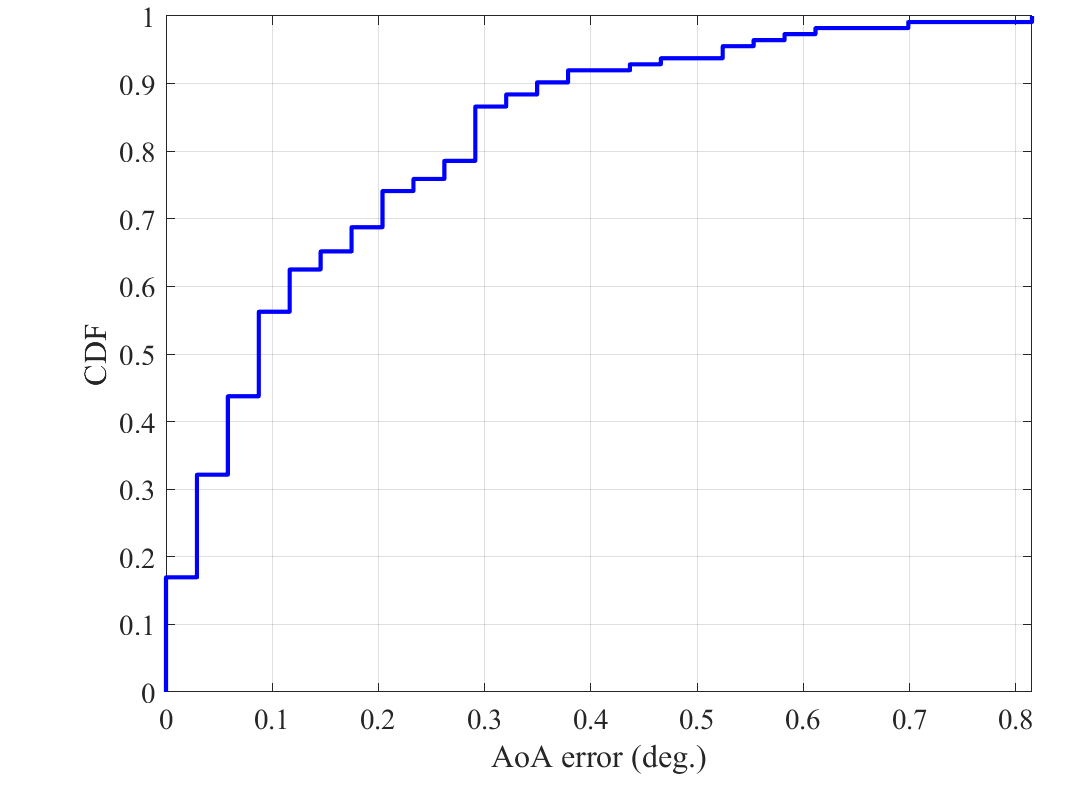}
    \caption{Empirical CDFs of path-parameter estimation errors from the
    controlled validation experiment: (a) delay error and (b) AoA error.}
    \label{fig:path_validation_cdf}
\end{figure}

The path-parameter extraction accuracy was evaluated using a controlled
experiment with known geometry. The receiver was mounted on a rotator and
measured at 120 control points, producing known changes in path length and
AoA. The extracted delay and AoA were then compared against the
corresponding values computed from the known geometry.
Fig.~\ref{fig:path_validation_cdf} shows the resulting empirical CDFs.
The average errors are $0.1$~ns in delay and $0.2^\circ$ in AoA.

\balance
\bibliographystyle{IEEEtran}
\bibliography{main_revised}

\end{document}